\documentclass[a4paper,11pt]{article}

\usepackage{graphicx,array}

\usepackage{color}
\usepackage{latexsym}
\usepackage{amsthm}
\usepackage{amsmath}
\usepackage{amssymb}
\usepackage{multirow}
\usepackage{empheq}
\usepackage{siunitx}
\usepackage{multicol}
\usepackage{dsfont}
\usepackage{epsfig}
\usepackage{slashed}
\usepackage{bbold}
\usepackage{psfrag}
\usepackage[svgnames]{xcolor}
\usepackage{easyReview}
\usepackage{subcaption}
\usepackage{xfrac}
\usepackage{multirow}
\usepackage{booktabs}
\usepackage{transparent}
\usepackage{soul}
\usepackage[normalem]{ulem}
\usepackage{jheppub} % for details on the use of the package, please see the JHEP-author-manual
\usepackage{tcolorbox} 
\usepackage{booktabs}  
\usepackage{caption}  
\usepackage{array}   
\usepackage{colortbl} 
\definecolor{pinkred}{RGB}{255, 105, 180} % Colore rosso tendente al rosa
\sethlcolor{pinkred}
\newtcolorbox{highlightbox}{colback=pinkred!20, colframe=pinkred, arc=5pt, auto outer arc}
\newcommand{\be}{\begin{equation}}
\newcommand{\ee}{\end{equation}}
\newcommand{\bea}{\begin{eqnarray}}
\newcommand{\eea}{\end{eqnarray}}
\newcommand{\bei}{\begin{itemize}}
\newcommand{\eei}{\end{itemize}}

\newcommand{\lag}{{\cal L}}

\definecolor{nicegreen}{rgb}{0.1,0.5,0.1}

\title{
Scalar Rayleigh  Dark Matter: current bounds and future prospects}
\author[a, b]{Daniele Barducci,}
\author[b]{Dario Buttazzo,}
\author[a, b, e]{Alessandro Dondarini,}
\author[c]{Roberto Franceschini,}
\author[a, b]{Giulio Marino,}
\author[d]{Federico Mescia,}
\author[a, b]{Paolo Panci}

\affiliation[a]{Dipartimento di Fisica ``Enrico Fermi", Universit\`a di Pisa, Largo Bruno Pontecorvo 3, I-56127, Pisa, Italy}
\affiliation[b]{INFN, Sezione di Pisa, Largo Bruno Pontecorvo 3, I-56127 Pisa, Italy}
\affiliation[c]{Universit\`a degli Studi Roma Tre and INFN, Via della Vasca Navale 84, I-00146, Rome, Italy}
\affiliation[d]{INFN, Laboratori Nazionali di Frascati, Via Enrico Fermi 54, I-00044 Frascati, Italy}
\affiliation[e]{Galileo Galilei Institute for Theoretical Physics, Largo Enrico Fermi 2, I-50125 Firenze,
Italy}
\emailAdd{daniele.barducci@pi.infn.it}
\emailAdd{dario.buttazzo@pi.infn.it}
\emailAdd{alessandro.dondarini@phd.unipi.it}
\emailAdd{roberto.franceschini@uniroma3.it}
\emailAdd{giulio.marino@phd.unipi.it}
\emailAdd{federico.mescia@lnf.infn.it}
\emailAdd{paolo.panci@unipi.it}

\abstract{
Dark Matter can interact with electroweak gauge bosons via higher-dimensional operators, in spite of being neutral under gauge interactions, much like neutral atoms interact with photons through Rayleigh scattering. This study explores effective interactions between a real scalar Dark Matter particle, singlet under the SM gauge group, and electroweak gauge bosons. We present a comprehensive analysis of current constraints and projected sensitivities from both lepton and hadron colliders as well as direct and indirect detection experiments in testing Rayleigh Dark Matter interactions. 
We find that, thanks to the complementarity between collider experiments and cosmological probes, thermally produced Rayleigh Dark Matter at the hundreds of GeV scale can be thoroughly tested with the next generation of experiments. For lighter candidates, upcoming forecasts will explore uncharted parameter space, significantly surpassing the thermal Dark Matter benchmark.}

\begin{document} 

\maketitle

\section{Introduction}\label{sec:intro}
Understanding the origin and nature of Dark Matter (DM) remains one of the foremost challenges in modern physics~\cite{Cirelli:2024ssz,Bertone:2016}. An extensive research effort is currently underway to detect DM through various approaches, including direct detection~\cite{Undagoitia_2016}, indirect detection~\cite{ID} and collider experiments~\cite{ColliderSearch}. Although DM should be neutral under electromagnetism, it can still interact with Standard Model (SM) photons and other electroweak (EW) gauge bosons via higher-dimensional operators, once heavy beyond the Standard Model (BSM) degrees of freedom are integrated out.
Interestingly, only a limited number of operators, the so-called Rayleigh ones, see {\it e.g.}\,\cite{Rajaraman:2012fu,Frandsen:2012db,Weiner:2012cb,Weiner:2012gm,
Liu:2013gba,Crivellin:2014gpa,
Crivellin:2015wva,Latimer:2016kdg,Kavanagh:2018xeh,Arina:2020mxo,Jackson_2013,Godbole_2015}, can produce these interactions at leading order in the Effective Field Theory (EFT) expansion. As this type of DM does not couple with quarks and gluons, the Rayleigh scattering cross-section is loop-suppressed and thus very challenging to unveil throughout direct detection (DD) experiments. On the other hand, it is a promising candidate for indirect detection (ID) probes. For example, the FERMI satellite can offer valuable insights by testing DM annihilation into photons up to about 500~GeV in DM mass.

In this context, we examine how this EFT scenario for scalar DM can be tested at near-future and next-generation colliders and how direct and indirect detection experiments 
might provide additional information about the model. Specifically, we consider current bounds stemming from direct DM searches at large hadron collider (LHC) and corresponding projections for its high luminosity phase (HL-LHC). Moreover, we make a survey on the reach that can be attained by the various proposed projects of the next generation of  experiments at colliders.
Among these we consider 
the Future Circular Collider (FCC) to be hosted at CERN in both proton-proton\,\cite{FCC:2018vvp} (FCC-hh) and electron-positron\,\cite{TLEPDesignStudyWorkingGroup:2013myl,FCC:2018byv,FCC:2018evy,Blondel:2019yqr} (FCC-ee) modes as well as the CEPC\,\cite{CEPCStudyGroup:2018rmc,CEPCStudyGroup:2018ghi} electron-positron Higgs factory planned to be build in China. Finally, we also discuss the reach of a high-energy
 muon collider\,\cite{AlAli:2021let,Aime:2022flm,Accettura:2023ked} ($\mu$C) operating at multi-TeV energies. As regarding direct and indirect DM searches we update the constraints from LZ-2022\,\cite{PhysRevLett.131.041002}, XLZD\,\cite{XLZD2022}, FERMI\,\cite{Fermi-LAT:2009ihh,Fermi-LAT:2015att} and CTA\,\cite{CTAConsortium:2010umy,CTA:2020qlo}.  

 Our results for all these probes of DM are summarized in Fig.\,\ref{fig:fig_sc_finalplot} and Fig.\,\ref{fig:fig_wc_finalplot} for two different types of UV completions for the EFT that we introduce below in Sec.~\ref{sec:framework}.
 This comprehensive comparison between current and future experiments aims to clarify how this EFT for DM can be tested in the near and more distant future through the interplay of multiple probes. It also serves as a guideline to assess the advantages and disadvantages  of the various collider options being considered for the post-LHC era.

The rest of the paper is organized as follows. In Sec.\,\ref{sec:framework} we describe the EFT framework that we use for our analysis. In Sec.\,\ref{sec:colliders} we analyze its collider phenomenology, while in Sec.\,\ref{sec:relic}, Sec.\,\ref{sec:DD} and Sec.\,\ref{sec:astro}
we discuss thermal freeze-out production mechanism and both direct and indirect detection phenomenology, respectively. Our results are then collected in Sec.\,\ref{sec:results} and we then summarize in Sec.\,\ref{sec:concl}. Finally, in App.\,\ref{app:UV} \ we discuss two possible ultraviolet (UV) completion for the Rayleigh operators.

\section{Theoretical framework}\label{sec:framework}

In this section, we re-examine the classification of effective DM-photon interactions outlined in\,\cite{Kavanagh:2018xeh}. This EFT framework assumes the existence of additional high-energy physics decoupled at an arbitrary scale $\Lambda$ which, in the low-energy limit, leaves only the light SM particles and a DM candidate with mass near the EW scale $v$. Our focus will be on the effective interactions involving a real  scalar DM candidate $\phi$, singlet under the SM gauge group, that interacts with the SM EW bosons through $d=6$ irrelevant operators\,\footnote{Analogous operators can be constructed for Majorana and Dirac fermionic DM. These operators present however an additional power suppression in terms of the scale $\Lambda$, making their effect smaller with respect to the scalar DM case. Moreover, Dirac fermions can feature a $d=5$ dipole moment operator with the hypercharge weak boson, which is tightly constrained. For this reason, in this work we only focus on the scalar DM case.}.
Specifically, we analyze the following $d=6$ operators
\begin{eqnarray}
{\cal L}_\phi = 
\frac{1}{2}\frac{g_\textsc{Y}^2}{16 \pi^2}
{\cal C}_{{\cal B}} \phi^2 B_{\mu\nu}B^{\mu\nu} + 
\frac{1}{2}\frac{g_\textsc{w}^2}{16 \pi^2}
{\cal C}_{{\cal W}} \phi^2 W^a_{\mu\nu}W^{a,\mu\nu} \label{eq:ops_1} \, ,
\end{eqnarray}
where $W_{\mu\nu}^a$ and $B_{\mu\nu}$ are the $SU(2)_L$ and $U(1)_Y$ field strength tensors  with $a=1,2,3$ and 
$g_{\textsc{w,y}}$ the corresponding gauge couplings.
In writing Eq.\,\eqref{eq:ops_1} we have absorbed the new physics scale $\Lambda$ in the definition of the Wilson coefficients ${\cal C}_{{\cal B},{\cal W}}$, which therefore have 
mass dimension of ${\rm GeV}^{-2}$. 
We have also normalized the effective operators of Eq.\,\eqref{eq:ops_1} with an explicit $g_{\textsc{w,y}}/(16\pi^2)$ loop factor, since operators involving a field strength tensor can arise only at loop-level in weakly coupled UV completions involving scalar, fermion and vector degrees of freedom~\cite{Buchmuller:1985jz,Craig:2019wmo}. Thus, Eq.\,\eqref{eq:ops_1} represents the EFT description under the assumption of a loop-level UV completion whose new physics states have mass equal to $\Lambda$.
In this case the Wilson coefficients ${{\cal{C}}_{{\cal B},{\cal W}}}$ are related to the scale where the corresponding new physics states can be integrated out, leading to a contact-interaction
\begin{eqnarray}
\frac{1}{\sqrt{{{\cal{C}}_{{\cal B},{\cal W}}}} } \equiv \frac{\Lambda_{\rm loop}}{g_{\rm loop}}  \ ,
\label{eq:weak_matching}
\end{eqnarray}
where by $g_{\rm loop}$ we denote a generic coupling characteristic of the new physics sector.

In other UV completions, however, the same operators can  be generated at tree-level, for instance by the exchange of a spin$-2$ particle with suitable couplings. Therefore, in order to be agnostic with respect to the UV origin of the effective operators, and also to facilitate
comparison with existing literature, we also map our results in what we dub {\emph{tree level normalized}} scenario, which we define by rescaling the Wilson coefficients
${\cal C}_{{\cal B},{\cal W}}$ by the loop factor and the corresponding gauge coupling. We then define
\begin{eqnarray}
{\cal L}^{\rm tree}_\phi = 
{\tilde{\cal C}}_{{\cal B}} \phi^2 B_{\mu\nu}B^{\mu\nu} + 
{\tilde{{\cal C}}}_{{\cal W}} \phi^2 W^a_{\mu\nu}W^{a,\mu\nu} \label{eq:ops_1_strong} \, .
\end{eqnarray}
In this case the Wilson coefficients ${\tilde{{{\cal{C}}}}_{{\cal B},{\cal W}}}$ are related to the 
 scale where these operators are generated as
\begin{eqnarray}\label{eq:scales}
\frac{1}{ \sqrt{{\tilde{{{\cal{C}}}}_{{\cal B},{\cal W}}}} } \equiv \frac{1}{g_{\textsc{y,w}}} \frac{\Lambda_{\rm tree}}{g_{\rm tree}} = \frac{4\pi \sqrt 2}{g_{\textsc{y},\textsc{w}}} \frac{\Lambda_{\rm loop}}{g_{\rm loop}} \simeq O(25-50) \frac{\Lambda_{\rm loop} }{g_{\rm loop}}\, ,
\label{eq:strong_matching}
\end{eqnarray}
where by 
%$g_{\rm loop}$ and 
$g_{\rm tree}$ we denote the unknown couplings of the new physics states coupled at tree-level with the SM and DM sector, taking into account possible SM couplings in the UV model by adopting a similar choice as for the loop level hypotheses.  
We discuss such details of possible UV completions in App.\,\ref{app:UV}, highlighting the range of applicability of the obtained results. We emphasize that throughout this work all bounds are presented in terms of the tree-level normalized new physics scale $\Lambda_{\rm tree}$ as defined in Eq.\,\eqref{eq:strong_matching}.
 The EFT coupling of DM with EW gauge bosons becomes clearer for energies below the EW scale $v$. In the SM broken phase, the operators in Eq.\,\eqref{eq:ops_1} can be projected onto the basis of physical gauge bosons as
\begin{eqnarray}
 & {\cal L}_\phi = 
 \dfrac{1}{2}\dfrac{e^2}{16 \pi^2}\phi^2\left(
{\cal C}_{\gamma\gamma}
A_{\mu\nu}A^{\mu\nu} 
+
{\cal C}_{\rm ZZ}  Z_{\mu\nu}Z^{\mu\nu}  +  {\cal C}_{\gamma \rm Z}
   Z_{\mu\nu}A^{\mu\nu} +  
{\cal C}_{\rm WW} W_{\mu\nu}^+ W^{-,\mu\nu}
 \right) \,\label{eq:ops_broken_1}\,.
\end{eqnarray}
Here $A_{\mu\nu}$, $Z_{\mu\nu}$, and $W_{\mu\nu}^\pm$ denote the field strength tensors for the photon, $Z$ boson, and $W^{\pm}$ boson, respectively. The electric charge is given by $e = g_Y c_\text{w} = g_\text{w} s_\text{w}$, where $c_{\rm w}$ and $s_{\rm w}$ represent the cosine and sine of the Weinberg angle, respectively and
\begin{equation}
\begin{alignedat}{2}
    & {\cal C}_{\gamma\gamma} = {\cal C}_{{\cal B}} + {\cal C}_{{\cal W}}, 
    &\quad {\tilde{\cal C}}_{\gamma\gamma} = {\tilde{\cal C}}_{{\cal B}} c_{\rm w}^2 + {\tilde{\cal C}}_{{\cal W}} s_{\rm w}^2, \\
    & {\cal C}_{\rm ZZ} = {\cal C}_{{\cal B}} \tan^2_{\rm w} + {\cal C}_{{\cal W}} \cot^2_{\rm w}, 
    &\quad {\tilde{\cal C}}_{\rm ZZ} = {\tilde{\cal C}}_{{\cal B}} s_{\rm w}^2 + {\tilde{\cal C}}_{{\cal W}} c_{\rm w}^2, \\
    & {\cal C}_{\gamma \rm Z} = -2 \big({\cal C}_{{\cal B}} \tan_{\rm w} - {\cal C}_{{\cal W}} \cot_{\rm w}\big), 
    &\quad {\tilde{\cal C}}_{\gamma \rm Z} = - \big({\tilde{\cal C}}_{{\cal B}} - {\tilde{\cal C}}_{{\cal W}}\big) s_{\rm 2w}, \\
    & {\cal C}_{\rm WW} = \frac{2}{s^2_{\rm w}} {\cal C}_{{\cal W}}, 
    &\quad {\tilde{\cal C}}_{\rm WW}= 2 {\tilde{\cal C}}_{{\cal W}}.
\end{alignedat}
\label{eq:EWSB-coeff}
\end{equation}

\begin{figure}[t!]
    \centering
    \includegraphics[width=0.45\linewidth]{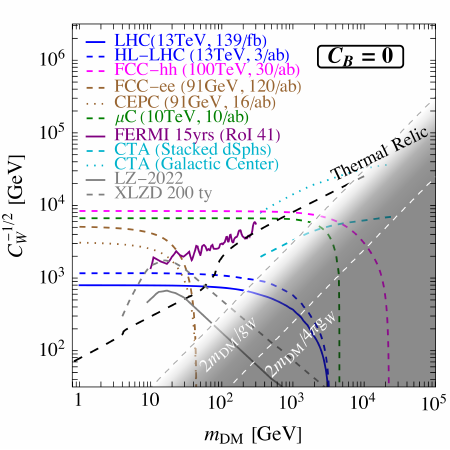}\hspace{0.5cm}
    \includegraphics[width=0.45\linewidth]{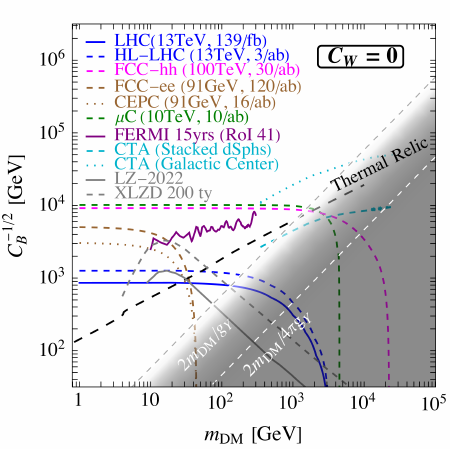}
    \caption{Bounds on ${\tilde {\cal C}}^{-1/2}_{\mathcal{B,W}}$, as defined in Eq.~\eqref{eq:strong_matching}, for a scalar Rayleigh DM candidate in the ${\tilde {\cal C}}_{\cal B} = 0$ (\textit{Left panel}) and ${\tilde {\cal C}}_{\cal W} = 0$ (\textit{Right panel}) planes. Solid lines represent current experimental constraints, while dashed lines illustrate future projections. The solid blue line represents the bounds from the \textsc{LHC}, while the dashed blue line corresponds to the high-luminosity phase projection. Brown, magenta, and green dashed lines indicate the projections for FCC-ee at the $Z$-pole, FCC-hh and the $\mu$C, respectively. Regarding LHC, HL-LHC, FCC-hh, $\mu$C the lines (both solid and dashed) represent the most stringent constraints obtained from the DY and VBS searches.
    Additionally, the dashed black line marks the thermal relic target, and the solid purple line indicates the bound from the \textsc{FERMI} experiment. Gray lines represent direct detection constraints, with the solid line for LZ-2022 and the dashed line for the projected next-generation \textsc{XLZD} experiment. The diagonal gray gradient represents the region allowed by the requirement $\Lambda_{\rm tree} > 2m_{\rm DM}$. The darker the region, the closer the coupling $g_{\rm tree}$ approaches its maximum value allowed by perturbative unitarity, $g_{\rm tree} \sim 4\pi$.}
    \label{fig:fig_sc_finalplot}
\end{figure}

For all the probes considered in this work — collider, direct detection, and indirect detection — it is crucial to ensure the validity of the EFT. To this end, we impose that the new physics scale $\Lambda$ probed by the experiment is such that 
\begin{equation}
    \Lambda > q > 2 m_{\rm DM} \ ,
    \label{eq:EFTcondition}
\end{equation}
where $q$ generally represents the momentum transferred, while for annihilation processes it represents the effective $\sqrt{s}$.
As explicitly shown in Eq.~\eqref{eq:weak_matching} 
and Eq.~\eqref{eq:strong_matching}, this condition depends on the values of the UV couplings \(g_{\rm tree}\) and \(g_{\rm loop}\). However, these couplings are constrained by the perturbative unitarity limit, with their maximum value restricted to \(4\pi\). To illustrate this, in Fig.~\ref{fig:fig_sc_finalplot}, we show a gray gradient-shaded region where darker shades correspond to couplings closer to the perturbative unitarity limit.

\section{Colliders}\label{sec:colliders}

In this section we discuss how present and future colliders can constrain the Wilson coefficients of the Rayleigh operators defined by Eq.\,\eqref{eq:ops_1}, or equivalently by Eq.\,\eqref{eq:ops_1_strong}. 
We consider in particular the LHC  running at $\sqrt s=13\,$TeV at the current phase ($139\,\si{fb}^{-1}$) and until the end of the high-luminosity phase when $\simeq\,3\,$ab$^{-1}$ of integrated luminosity will have been collected,
 as well as the future hadron collider FCC-hh operating at $\sqrt s=80\,$TeV and  $\sqrt s=100\,$TeV\,\cite{FCC:2018vvp}. We also consider $Z$ factories such as FCC-ee\,\cite{TLEPDesignStudyWorkingGroup:2013myl,FCC:2018byv,FCC:2018evy,Blondel:2019yqr} and CEPC\,\cite{CEPCStudyGroup:2018rmc,CEPCStudyGroup:2018ghi} and a $\mu$C operating at multi-TeV energies\,\cite{AlAli:2021let,Aime:2022flm,Accettura:2023ked}. In order to perform our analysis we have implemented the interactions of Eq.\,\eqref{eq:ops_1} 
into the {\tt FeynRules}\,\cite{Alloul:2013bka} package using the {\tt UFO}\,\cite{Degrande:2011ua} format to compute cross-section and generate Monte Carlo events\footnote{For hadron colliders the {\tt MadGraph5\_aMC@NLO} default parton distribution function NNPDF2.3\,\cite{Ball:2013hta} has been employed.} with {\tt MadGraph5\_aMC@NLO}\,\cite{Alwall:2014hca}.
The results are then collectively presented in Sec.\,\ref{sec:results} together with the limits arising from complementary constraints.
For the collider analyses, in order to compute the experimental selection acceptances on the signal and background rates
and extract the bounds on the Wilson coefficients of Eq.\,\eqref{eq:ops_1} 
we rely on leading order (LO) parton level simulations. 
Regarding future lepton colliders such as FCC-ee, CEPC, and the $\mu$C, this can be considered a good approximation given the exploratory nature of this study, the generally smaller higher-order EW corrections compared to QCD ones, and the expected cleanliness of the experimental environments. When recasting current LHC analyses, see Sec.\,\ref{sec:LHC}, we directly rely on the experimental results 
for the upper limits on the new physics cross-sections, which have been computed by the experimental collaborations with a proper treatment of the background yields and detector effects. 
By working at the parton level for the signal event rates, we neglect reconstruction and isolation efficiencies for detector-level objects, as well as parton showering effects, which might reduce the signal acceptance and thus weaken the limits compared to parton-level results.  
However, as we will discuss in Sec.\,\ref{sec:LHC_DY} and Sec.\,\ref{sec:LHC_VBS}, full-fledged detector simulations indicate that these changes in acceptance will modify the bounds on the NP scales only by ${\cal O}(10\%-30\%)$, depending on the specific process under consideration. We consider these modifications small in the context of this work, which aims to provide a global perspective on the sensitivity of various colliders and experiments. Thus, we present our results at the parton level for both present and future colliders.  
Nonetheless, we caution the reader that our results may be subject to small modifications and should be interpreted with this in mind.

For all the future collider studies the bounds that we show are 95\% confidence level (CL) limits, computed by imposing a value $z=2$ for the statistical significance $z$ defined as 
\begin{equation}
\label{eq:significance}
    z = \frac{N_S}{\sqrt{N_B}} \ ,
\end{equation}
where $N_{S,B}$ are the number of signal and background events respectively. This definition that we adopt only takes into account statistical uncertainties on the background determination, and therefore our results represent the optimal reach that can be obtained with the proposed analyses, which will be slightly  degraded by the inclusion of systematic uncertainties.

\subsection{LHC and high-luminosity LHC}\label{sec:LHC}

\subsubsection{Drell-Yan production}\label{sec:LHC_DY}

\begin{figure}[t!]
    \centering
    \includegraphics[width=0.99\linewidth]{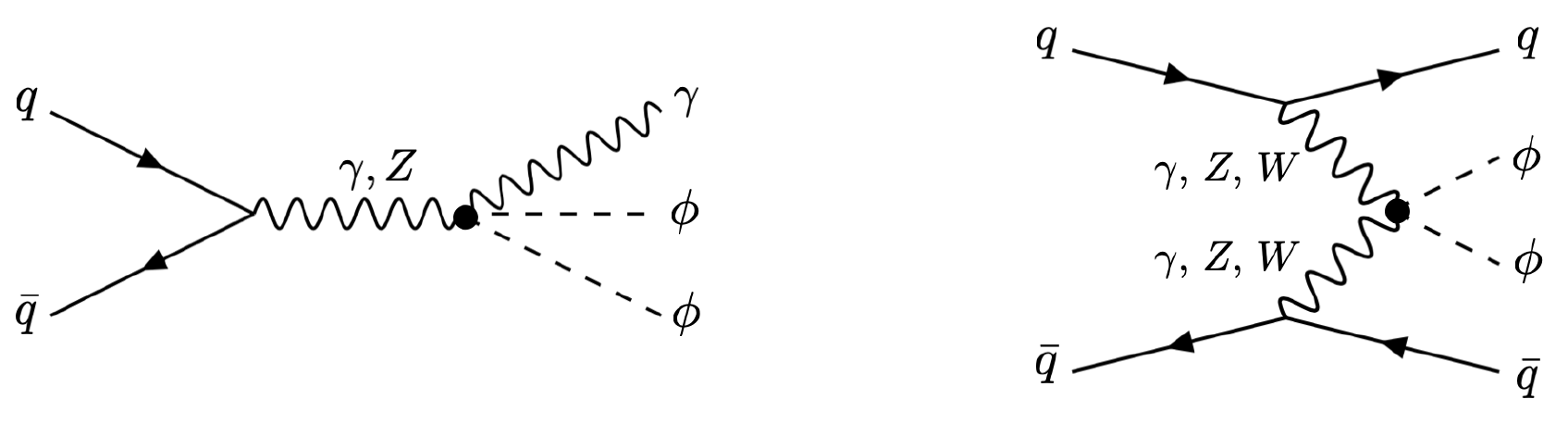}
    \caption{
    Representative Feynman diagrams for DY mono$-\gamma$ pair-production (left) and VBS production (right) at an hadron collider of scalar $\phi$ particles through the operators of Eq.\,\eqref{eq:ops_1}. The black dot represents an insertion of the effective operator.}
    \label{fig:LHC_DY}
\end{figure}

We compute current LHC bounds from Drell-Yan (DY) production by recasting the ATLAS search
for DM in association with a high energetic photon
performed at $\sqrt s=13\,$TeV with an integrated luminosity of 139\,fb$^{-1}$\,\cite{ATLAS:2020uiq}.
We consider the DY signal process 
\begin{equation}
p p \to \gamma \phi \phi
\end{equation}
whose Feynman diagram is shown in the left panel of Fig.\,\ref{fig:LHC_DY}. We generate LO parton level events and we apply the event selections of the ATLAS search in order to estimate the event yield in each analysis signal region.
At  analysis level a baseline selection of photon candidates fixes 
\begin{equation}
E_T^\gamma>150\,\si{GeV}\text{, }|\eta_\gamma|<1.37\text{ or }1.52<|\eta_\gamma|<2.37\,.
\end{equation}
ATLAS defines
seven signal regions with different inclusive or exclusive requirements on $E_T^{\rm miss}$.
We find that the most constraining signal region is an ``exclusive'' signal region, following the naming scheme from ATLAS, defined by the requirement $E_T^{\rm miss}>375\,$GeV. 
In this phase space region the signal cross-section for $m_\phi=5~$GeV and ${\tilde{{\cal C}}}_{{\cal W}}=0$ reads $\sigma_{\rm sig}\simeq 0.27\,\left(\frac{{\tilde{{\cal C}}}_{{\cal B}}}{10^{-6}\,{\rm GeV}^{-2}}\right)^2\,{\rm fb}$.

To estimate the impact that reconstuction and isolation efficiencies as well as parton showering effects have on the bounds we have derived, we have performed a full-fledged recast of the DY ATLAS search for the case $m_{\rm DM}=0$ and ${\tilde{\cal C}}_{\cal W}=0$. Specifically hadronisation and decay of unstable particles have been performed through {\tt pythia8}\,\cite{Sjostrand:2007gs} while
            {\tt Delphes3}\,\cite{deFavereau:2013fsa} has been employed for a fast detector simulation.            Jets have been reconstructed with {\tt FastJet}\,\cite{Cacciari:2011ma} via the anti$-\kappa_T$ algorithm\,\cite{Cacciari:2008gp}. Detector simulation
		has been implemented via {\tt Delphes3}, relying on the default ATLAS {\tt delphes\_card}, while analysis level cuts have been applied using {\tt MadAnalysis5}\,\cite{Conte:2012fm}.
        After this procedure, we obtain that the most 
        constraining  signal region is still the one with $E_T^{\rm miss}>375\,$GeV and that the ratio of the parton level over the full fledged detector level simulation acceptances is $\sim 1.5$, which reflects on a loosening of the bound on the new physics scale $\Lambda$ of a factor $\sim1.5^{-1/4}\sim 10\%$, which we deem as small with respect to the goal of this study and we will therefore present our results working at the parton level.

As LHC can produce scatterings with exchanged momentum higher than the scale of the effective operators, we need to make provisions to check that our EFT description be valid. The spirit of this check is to ensure that the bound we obtain is not influenced by events with momentum exchanged larger than the mass scale of the new states of the UV completion which give rise to the Rayleigh operator once integrated out.
A necessary condition for the validity of the EFT description is  the cut-off scale $\Lambda_{\rm loop}$ as defined in Eq.\,\eqref{eq:weak_matching} not be lower than the $E_T^{\rm miss}$ cut applied in the analysis. As the momentum exchanged in the effective vertex in the DY processes of Fig.\,\ref{fig:LHC_DY} is of order $E_T^{\rm miss}$, when
\bea E_T^{\rm miss} > \Lambda_{\rm loop} \eea
 one expects to see the effects of the new physics particles from the loop-level UV completions, which may chance the rates for the signal compared to the EFT prediction. 
We find that  significant deviations from the EFT predictions are to be expected at  the LHC for the size of $\Lambda_{\rm loop} $ that  can be probed. This removes interest from an EFT description and invites to look for direct evidence of the new states in the UV completion. The LHC, instead, can put meaningful bounds on models with tree-level mediators, which are suitable for the description of Eq.~\eqref{eq:strong_matching}.

 \begin{table}[t!]
    \centering
    \rowcolors{1}{gray!10!white}{white} % Alterna i colori delle righe
    \begin{tabular}{ c @{\hskip 10pt} | @{\hskip 10pt} c @{\hskip 10pt} | @{\hskip 10pt} c }
    \toprule
    \rowcolor{white!10} 
    \multicolumn{3}{c}{mono-$\gamma$ DY,~~$\sqrt{s}=13\,$TeV ~~${\tilde{{\cal C}}}_{\cal W}=0$} \\ 
    %\rowcolor{lime!20}
    \multicolumn{3}{c}{LHC~~$({\cal L}=139\,$fb$^{-1})$~~and HL-LHC~~$({\cal L}=3\,$ab$^{-1})$} \\
    \midrule
    %\rowcolor{red!30!orange!50}
    %\rowcolor{blue!30!green!30}
    $m_{\rm DM}\,$[GeV] & ${\tilde {\cal C}^{-1/2}_{\cal B}} \,$[GeV] & ${\tilde {\cal C}^{-1/2}_{\cal B, \rm{HL}}} \,$[GeV] \\
    \midrule
    \rowcolor{white!10} 
    50    & 842   & 1236  \\ 
    \rowcolor{white!10} 
    100   & 808   & 1186  \\ 
    \rowcolor{white!10} 
    500   & 534   & 784   \\  
    \rowcolor{red!20}  
    1000  & 307   & 451   \\ 
    \rowcolor{red!20}  
    1200  & 255   & 375   \\ 
    \rowcolor{red!20}  
    1500  & 178   & 261   \\ 
    \bottomrule
    \end{tabular}
    \caption{The bounds on the scale ${\cal \tilde C}^{-1/2}_{\cal B}$, assuming ${\cal \tilde C}_{\cal W}=0$, are derived from mono-$\gamma$ Drell-Yan processes at the current LHC run and projected for the high-luminosity phase, considering various DM masses.  We have highlighted in red the rows where the breakdown of the EFT occurs for both the LHC and HL-LHC. Specifically, this happens when $ g_{\rm tree} \sim 4\pi $ and hence $\tilde{\mathcal{C}}^{-1/2}_{\mathcal{B}} \sim 2m_{\rm DM}/4\pi g_{\textsc{Y}}$, thus for DM masses above 1 TeV.
    }
    \label{tab:LHC_results}
    \end{table}
In Tab.\,\ref{tab:LHC_results} we show the bounds for models with tree-level mediator, fixing ${\tilde C}_{\cal W}=0$ to ease the display of the result.
Under this assumption of coupling structure, when recasting the ATLAS analysis 
 we check \textit{a posteriori} that condition~\eqref{eq:EFTcondition} is satisfied.
Building on the  recast of the ATLAS search, we project the bounds achievable by a similar analysis at the end of the LHC high-luminosity phase with an integrated luminosity of $\simeq 3$ ab$^{-1}$.
We focus on the most constraining signal region of the current search, in which $E_T^{\rm miss} > 375$~GeV. We rescale the expected number of background events by the ratio of the target HL-LHC luminosity to that of the ATLAS analysis, and use Eq.~\eqref{eq:significance} to compute the statistical significance. In this case we find that masses up to 1200~GeV can be probed at the  HL-LHC consistently with a tree-level mediator EFT description.

\begin{figure}[t!]
    \centering
    \includegraphics[width=0.47\linewidth]{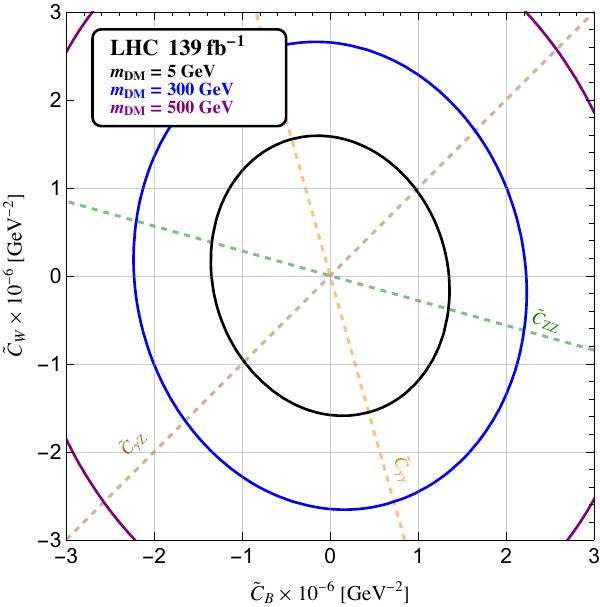}\hspace{0.5cm}
    \includegraphics[width=0.47\linewidth]{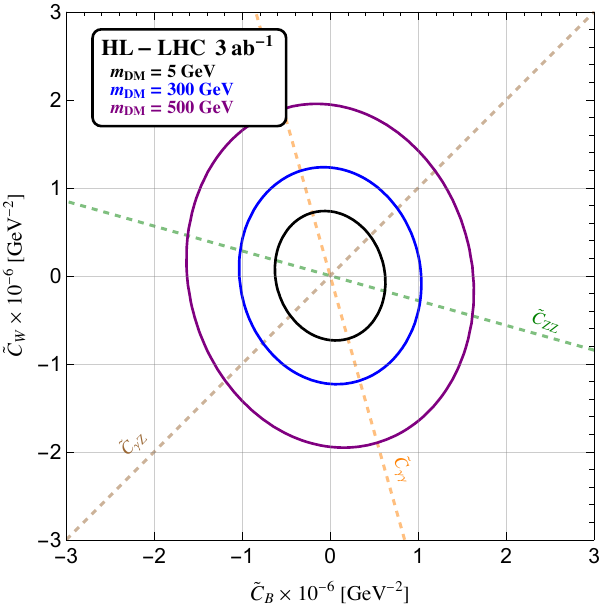} \\
\caption{Constraints in the $({\tilde {\cal C}}_{\cal B},{\tilde {\cal C}}_{\cal W})$ plane for DY process at the LHC (\textit{left panel}) and the high-luminosity LHC (\textit{right panel}), shown for various DM mass values. The dashed lines indicate the directions associated with the couplings defined in Eq.~\eqref{eq:EWSB-coeff}. The orange, green, and brown lines correspond to the directions where ${\tilde {\cal C}}_{\gamma\gamma}$, ${\tilde {\cal C}}_{\rm ZZ}$, and ${\tilde {\cal C}}_{\gamma \rm Z}$ vanish, respectively.}

    \label{fig:LHC-planes}
\end{figure}

The sensitivity of the LHC on   the $({\tilde C}_{\cal B},{\tilde C}_{\cal W})$ plane is shown in Fig.~\ref{fig:LHC-planes} for both the LHC at $139\,{\rm fb}^{-1}$ and the high-luminosity phase at $3\,{\rm ab}^{-1}$. For a given $m_{\rm DM}$ mass, the regions outside the ellipses are excluded at $95\%$ CL. Additionally we show the directions along which ${\tilde {\cal C}}_{\gamma\gamma}$, ${\tilde {\cal C}}_{ZZ}$, and ${\tilde {\cal C}}_{\gamma Z}$ vanish. We observe that the HL-LHC, thanks to its ability to excite both high-energy $Z$ bosons and $\gamma$, has a broadly equivalent sensitivity in all the directions of the plane. 

\subsubsection{Vector-boson scattering production}\label{sec:LHC_VBS}

We extract current LHC bounds from searches of vector boson scattering (VBS) signatures by recasting the ATLAS search for an invisible decaying Higgs boson performed at $\sqrt{s}=13$~
TeV with 139\,fb$^{-1}$ of integrated luminosity\,\cite{ATLAS:2022yvh}. We compute the
VBS production 
\begin{equation}
p p \to j j \phi \phi
\end{equation}
Unlike for DY mono$-\gamma$ production, here also the contribution from $W$ boson scattering enters the signal rate. We compute the signal rates  
at LO in the perturbative expansion. We apply the selection of the ATLAS search\,\cite{ATLAS:2022yvh} on our partons. Events are selected by requiring the presence of at least two jets,  the leading and sub-leading one satisfying 
$$p_T^{j_{1,2}}>80\,{\rm GeV},\,50\,\text{ GeV with }|\eta_j|<4.5\,.$$ 
The two jets should be in opposite hemisphere $$\eta_{j_1}\eta_{j_2}<0\,$$ and have a large pseudorapidity separation, $$\Delta \eta_{j_1 j_2}>3.8\,.$$ Further, to suppress the large QCD background, they should not be back to back in the azimuthal plane, $$\Delta\phi_{j_1 j_2}<2$$ and the missing transverse energy of the event  should be $$E_T^{\rm miss} >  160~\text{GeV}\,.$$

In the ATLAS selection several signal regions are defined with increasingly stricter $m_{jj}$ requirements and further splitting in $\Delta \phi_{jj}$ and $E_T^{\rm miss}$ bins.
%, where $m_{jj}$ is the invariant mass of the di-jet system. 
We find that the most constraining signal region is the one with with the requirement 
\begin{equation} 
    E_T^{\rm miss}>200\,\text{GeV,  } \Delta \phi_{jj}<1,  \; m_{jj} \in [2,3.5]\,\text{ TeV} \ , \label{eq:VBF-SR}
\end{equation}
for which the signal cross-section for $m_\phi=0$ and ${\tilde{{\cal C}}}_{{\cal W}}=0$ reads $\sigma_{\rm sig}\simeq 0.063\,\left(\frac{{\tilde{{\cal C}}}_{{\cal B}}}{10^{-6}\,{\rm GeV}^{-2}}\right)^2\,{\rm fb}$.

Based on the above results we also estimate the reach of a similar analysis at the HL-LHC, considering an integrated luminosity 3~ab$^{-1}$. We estimate the sensitivity of HL-LHC taking into account only the impact of statistical uncertainties. 
As for the case of the DY study, we have performed again a full-fledged recast of the ATLAS VBS search for the case $m_{\rm DM}=0$ and ${\cal C}_{\cal W}=0$. In this case, due to the more involved final state with respect to the DY mono$-\gamma$ study, the bounds get degraded by a factor ${\cal O}(30\%)$. To be on a more conservative side, we have then decided to rescale the limits arising from the,
less stable under detect effects, VBS analysis by a flat factor 30\%. The same rescaling factor is applied to the projections for FCC-hh in Sec.\,\ref{sec:FCC-hh_VBF}.

Our results are shown in Fig.~\ref{fig:fig_sc_finalplot}, where we show the most stringent bound arising from DY and VBS processes.
The VBS process provides a significantly larger contribution in the $C_{\cal B} = 0$ plane, since a non-zero coupling to the $W$ boson enhances the contribution from the $WW \to \phi\phi$ sub-processes, leading to more stringent bounds. However, in both the $C_{\cal W} = 0$ and $C_{\cal B} = 0$ planes, the DY process remains the dominant production mechanism.

As for the DY-case, care must be paid in ensuring that the our results comply with the validity of the effective description. In VBS processes the momentum exchanged in the effective vertex  should be estimated by the transverse momentum of each DM particle. As this is not experimentally accessible, we prefer to use a proxy, that is half the invariant mass of the DM pair, $p_T^{\phi}\sim m_{{\rm \phi,\phi}}/2$. This quantity is not  experimentally measurable on each event, due to the unknown partonic center of mass energy entering the hard scattering process. However, it enjoys the property \(m_{{\rm \phi,\phi}} > 2 m_{{\rm DM}}\), and we can then verify \textit{a posteriori} the condition of Eq.~\eqref{eq:EFTcondition} for each DM mass hypothesized in the search to approximately ensure EFT validity.

\subsection{FCC-hh at \texorpdfstring{$\sqrt s=80\,$}{TEXT}TeV and \texorpdfstring{$\sqrt s=100\,$}{TEXT} TeV}\label{sec:FCC-hh}
Looking beyond the \textsc{LHC}, we consider a future hadron collider (FCC-hh) with a center-of-mass energy in the range $\SI{80}{\tera\electronvolt}$ to $\SI{100}{\tera\electronvolt}$. The collected integrated dataset for the experiments at this collider is expected to be in the $30\,$ab$^{-1}$ range\,\cite{Mangano:2022ukr}, which we use as benchmark target luminosity in the following for both center of mass energy options.

\subsubsection{Drell-Yan production}

\begin{table}[t!]
\begin{center}
\rowcolors{1}{gray!10!white}{white} % Alterna i colori delle righe
\begin{tabular}{ c | c c | c c }
\toprule
\rowcolor{white!10} 
\multicolumn{5}{c}{mono$-\gamma$ DY at FCC-hh~~$\sqrt{s}=80\,$TeV~~${\cal L}=30\,$ab$^{-1}$~~${\tilde{{\cal C}}}_{\cal W}=0$}\\
\hline
& \multicolumn{2}{c|}{{\bf w/o EFT validity}} & \multicolumn{2}{c}{{\bf with EFT validity}} \\
\hline
$m_{\rm DM}\,$[GeV] 
& $p_{T,{\rm min}}^\gamma\,$[GeV] 
& ${\tilde {\cal C}}^{-1/2}_{\cal B}\,$[GeV]
& $p_{T,{\rm min}}^\gamma\,$[GeV] 
& $p_{T,{\rm max}}^\gamma = {\tilde {\cal C}}^{-1/2}_{\cal B}\,$[GeV]  \\ 
\hline
\rowcolor{white!10} 
100 & 5500  & 7780 & 4000 & 7300 \\ 
\rowcolor{white!10} 
1000 & 6000  & 7350 & 4000 & 6650  \\ 
\rowcolor{white!10} 
2000 & 6500   & 6640 & 3500  & 5100 \\ 
\rowcolor{white!10} 
5000 & 8500  & 4490 & 200  & 250  \\ 
\bottomrule
\end{tabular}
\vskip 20pt
\begin{tabular}{ c | c c | c c }
\toprule
\rowcolor{white!10} 
\multicolumn{5}{c}{mono$-\gamma$ DY at FCC-hh~~$\sqrt{s}=100\,$TeV~~${\cal L}=30\,$ab$^{-1}$~~${\tilde{{\cal C}}}_{\cal W}=0$}\\
\hline
& \multicolumn{2}{c|}{{\bf w/o EFT validity}} & \multicolumn{2}{c}{{\bf with EFT validity}} \\
\hline
$m_{\rm DM}\,$[GeV] 
& $p_{T,{\rm min}}^\gamma\,$[GeV] 
& ${\tilde {\cal C}}^{-1/2}_{\cal B}\,$[GeV]
& $p_{T,{\rm min}}^\gamma\,$[GeV] 
& $p_{T,{\rm max}}^\gamma = {\tilde {\cal C}}^{-1/2}_{\cal B}\,$[GeV]  \\ 
\hline
\rowcolor{white!10} 
100 & 7000  & 9150 & 4500 & 8500 \\ 
\rowcolor{white!10} 
1000 & 7500  & 8800 & 5000 & 7900  \\ 
\rowcolor{white!10} 
2000 & 8000   & 8160 & 4500  & 6600 \\ 
\rowcolor{white!10} 
7000 & 11000  & 4850 & 300  & 380  \\ 
\bottomrule
\end{tabular}
\end{center}
\caption{
Bounds on ${\cal \tilde C}^{-1/2}_{B}$ with ${\cal \tilde C}_{W}=0$ from mono$-\gamma$ DY processes at FCC-hh with a center of mass energy of $\sqrt s=80\,$TeV (upper table) and $\sqrt s=100\,$TeV (lower table)  for various DM masses.
{In the first column, where the {\emph{EFT validity}} is not imposed, we report only the lower limit on optimal $p_T^\gamma$ value  requirement, while the columns {\emph{with EFT validity}} report both the lower and the upper limits, where the latter is equivalent to the bound on $\Lambda_{{\rm tree}}$.}
}
\label{tab:FCC-hh-optimal_cut}
\end{table}

We consider the mono$-\gamma$ DY signal processes of Fig.\,\ref{fig:LHC_DY}   and proceed in the following way for estimating the SM backgrounds. For the background estimation we leverage  the ATLAS 13\,TeV mono$-\gamma$ analysis\,\cite{ATLAS:2020uiq}. ATLAS finds the dominant SM background to be $p p \to Z\gamma,\, Z\to \nu \bar \nu$, followed by $pp\to W\gamma,\; W\to \ell \nu$. Other sources arise from  $Z(\to \ell \ell)\gamma$, $\gamma+j$, fake photons from $e$ and fake photons from $j$ which are sub-leading  and, the latter two, hard to estimate with fast  simulations.
The two dominant sources account for $\simeq 80\%$ of the total background yield, with the former contributing for a $\simeq\,60\%$. Moreover the ratio between the two is almost constant in all the ATLAS analysis signal regions of interest and this can be understood by the fact that, for the partonic center of mass energies $\hat s$ of interest for the ATLAS analysis (which are of the order of the $E_T^{\rm miss}$ cut) the relevant parton luminosities are in approximate constant ratio. 
Our {\tt MadGraph5\_aMC@NLO} calculation of the rate of this process in all the fiducial regions defined in terms of $E_T^{\rm miss}$ and $|\eta^\gamma|$ by the ATLAS analysis corresponds to $\simeq 80\%$ of the estimate from ATLAS for the same process. Therefore, we assume that missing higher order corrections and detector effects can be incorporated in our calculation by increasing the $Z\gamma$ rate obtained from {\tt MadGraph5\_aMC@NLO} by a constant factor 2 that makes its rate agree with the total background estimated by ATLAS.  

For an exploratory determination of the reach of FCC-hh we assume that this rescaling procedure, that holds well for LHC, remains valid at larger $\sqrt s$ and for tighter 
$E_T^{\rm miss}$ cuts.
This is motivated by the fact that a mono$-\gamma$ at higher center of mass energies $s$ is expected to be sensitive to comparable values of $\hat s/s$. Since working at constant $\hat s/s$ the ratio of the parton luminosity intervening in the $Z\gamma$ and $W\gamma$ is approximately constant, this justifies our rescaling assumption.

We thus estimate the total SM background at $\sqrt s=80\,$TeV and $\sqrt s=100\,$TeV by multiplying by a factor of two the rate of the dominant $Z\gamma$ background computed at LO with {\tt MadGraph5\_aMC@NLO}. 
We then compute the expected bounds on the Wilson coefficients of the operators of Eq.\,\eqref{eq:ops_1} by optimizing the cut on $E_T^{\rm miss}$
%, which again is equivalent to $p_T^\gamma$ working at LO parton level, 
while keeping fixed, for simplicity, the requirement $|\eta_\gamma|<2.37$ as in the 13\,TeV analysis. 
The signal, computed for $m_\phi=0$ and ${\tilde{{\cal C}}}_{{\cal W}}=0$, and background cross-sections for FCC-hh with $\sqrt s=80\,(100)\,$TeV  with $p_{T,{\rm min}}^{\rm \gamma}>5.5\,(7)\,$TeV are respectively $\sigma_{\rm sig}\simeq 1.55\,(2.24)\left(\frac{{\tilde{{\cal C}}}_{{\cal B}}}{10^{-6}\,{\rm GeV}^{-2}}\right)^2\,{\rm fb}$ and $\sigma_{\rm bkg}=1.35\times 10^{-3}\,(7.65\times 10^{-4})\,$fb.\footnote{Note that given the signal and background yields the use of the statistical significance of Eq.\,\eqref{eq:significance} is justified, in that a different choice as, {\it e.g.}, $z=N_S/\sqrt{N_S+N_B}$ yield almost the same results.}

\smallskip
In Tab.~\ref{tab:FCC-hh-optimal_cut} we show the optimal cuts and the resulting bounds on ${\cal \tilde C}^{-1/2}_{{\cal B}}$, as defined in Eq.\,\eqref{eq:strong_matching}, by fixing ${\cal \tilde C}_{\cal W}=0$.
In the columns labeled \emph{w/o EFT validity}, we report only the lower limit on 
$p_T^\gamma$, as setting an upper limit would be unnecessary if the EFT description 
is assumed to remain valid throughout. Conversely, in the columns {\emph{with EFT validity}}  we report both the lower and the upper limit on $p_T^\gamma$, where the latter is equivalent to the bound we find on $\Lambda_{{\rm tree}}$, given our requirement that no photon $p_T$ should exceed the physical scale $\Lambda$ where the new physics particles start to be propagating degrees of freedom. As in the case of the 13\,TeV LHC analyzed in Sec.\,\ref{sec:LHC}, we observe that, when requiring the validity of the EFT description, the bounds degrades at higher DM masses due to the shift of the $p_T^\gamma$ distribution towards harder values. To facilitate the comparison among the different analyses we report in Fig.\ref{fig:fig_sc_finalplot} the limits only imposing a lower limit on $p_T^\gamma$, estimating the EFT validity via Eq.\,\eqref{eq:EFTcondition}. 

\subsubsection{Vector-boson scattering production}\label{sec:FCC-hh_VBF}

For our estimate of the reach in the VBS channel at FCC-hh we leverage the background estimation of the 13\,TeV ATLAS search\,\cite{ATLAS:2022yvh}. In the signal region defined by Eq.~\eqref{eq:VBF-SR} the background yield 
 is dominated by EW $Z$ boson production,
which amount to approximately 1/3 of the total background rate. A comparable background rate arises from EW $W$  boson production, and the remainder of the background comes from $Z/W$   production in QCD processes and further multi-jet processes. 

For our background estimate at the FCC-hh we compute EW $Z$ production at LO with {\tt MadGraph5 aMC@NLO} in the high $m_{jj}$ region of the ATLAS search\,\cite{ATLAS:2022yvh} defined by Eq.~\eqref{eq:VBF-SR}, finding that our calculation overestimates the ATLAS results by approximately a factor 2. This is due to the lack of additional jet radiations in our calculation, which would be vetoed in the fiducial regions defined by ATLAS. 
We thus obtain a rescaling factor of 3/2, in order to match the the total background yield. 
To compute the bounds at higher $\sqrt s$ we proceed in the following way. We work at fixed $m_{jj}/s$,
keeping fixed all the remaining selection cuts of the 13\,TeV ATLAS analysis,
 and estimate the total SM background at $\sqrt s=80\,$TeV and $100\,$TeV by multiplying for the same factor the EW $Z$ production rate computed at LO, from which we compute the expected sensitivity again using Eq.\,\eqref{eq:significance} as for the HL-LHC projections.
The signal, computed for $m_\phi=0$ and ${\tilde{{\cal C}}}_{{\cal W}}=0$, and background cross-sections for FCC-hh with $\sqrt s=80\,(100)\,$TeV  with $m_{jj}>12.5\,(15)\,$TeV are respectively $\sigma_{\rm sig}\simeq 4.04\,(7.27)\left(\frac{{\tilde{{\cal C}}}_{{\cal B}}}{10^{-6}\,{\rm GeV}^{-2}}\right)^2\,{\rm fb}$ and $\sigma_{\rm bkg}=5.03\,(4.06)\,$fb.

As before, we choose the simple criterion in Eq.~\eqref{eq:EFTcondition} in order to ensure the validity of the effective description.

Results are shown in our summaries Figure~\ref{fig:fig_sc_finalplot}
and 
Figure~\ref{fig:fig_wc_finalplot}, in which the strongest between DY and VBF is displayed.
In App.\,\ref{sec:app_fcc} we discuss how the detector  design, and in particular its acceptance in $p_T$ and $\eta$, affect the reach of FCC-hh.

\subsection{FCC-ee and CEPC}

Other collider proposals that are discussed for further exploration after the LHC include $e^+ e^-$ facilities, running at center of mass energies ranging from the $Z-$pole, to the $t\bar t$ threshold. Among the various options being discussed, circular colliders such as the FCC-ee at CERN~\cite{TLEPDesignStudyWorkingGroup:2013myl,FCC:2018byv,FCC:2018evy,Blondel:2019yqr} and CEPC in China~\cite{CEPCStudyGroup:2018rmc,CEPCStudyGroup:2018ghi} stand out as high-intensity machines capable of furthering the exploration of the intensity frontier. High intensity exploration  at low energy is an ideal ground to apply EFT, as it guarantees the validity of the EFT picture and, thanks, to the high-intensity can deliver  sensitivity potentially better than that of LHC. Interestingly, both FCC-ee and CEPC plan to have a run at $\sqrt s=m_Z$, with a targeted integrated luminosity in the ${\cal O}(10-100)\,$ab$^{-1}$ range, yielding $O(10^{12})$ $Z$ bosons, which gained these machines the nickname of Tera$-Z$ factories.

\begin{table}[t!]
    \begin{center}
    \begin{tabular}{ c | c c | c }
    \toprule
    \multicolumn{4}{c}{DY at $e^+ e^-$~~$\sqrt{s}=m_Z$~~${\tilde{{\cal C}}}_{\cal W}=0$}\\
    \hline
    & \multicolumn{2}{c|}{
    ${\cal L}=16\,$ab$^{-1}$
    } & \multicolumn{1}{c}{${\cal L}=120\,$ab$^{-1}$} \\
    \hline
    \rowcolor{gray!10}
    $m_{\rm DM}\,$[GeV] 
    & $p_{T,{\rm min}}^\gamma$[GeV] 
    & ${\tilde {\cal C}}^{-1/2}_{\cal B}\,$[GeV]
    & ${\tilde {\cal C}}^{-1/2}_{\cal B}\,$[GeV] \\ 
    \hline
    1 & 37.5  & 3043 & 5036 \\ 
    10 & 32.5  & 2524  & 4176\\
    20 & 22.5  & 1715  & 2839 \\
    30 & 15  & 910  & 1505 \\
    %35 & 10  & 543  & 898 \\
    40 & 5  & 225  &373 \\
    \bottomrule
    \end{tabular}
    \end{center}
    \caption{
    Representative bounds on $\Lambda_{\rm tree} = {\cal {\tilde{C}}}^{-1/2}_{B}$ with ${\cal {\tilde{C}}}_{W}=0$ from DY processes at a $e^+ e^-$ colliders with  $\sqrt s=m_Z$ for various DM masses. The analysis is performed for both an integrated luminosity of $16$ab$^{-1}$ (CEPC) and $120$ab$^{-1}$ (FCC-ee).}
    \label{tab:cepc}
    \end{table}

Although these machines can run at higher energies, the drop in luminosity that it is entailed by going at higher energies in $e^+e^-$ collisions makes the Tera $Z$  run   the most worth being studied in detail. 
We   estimated  the sensitivity to the Rayleigh operators of Eq.\,\eqref{eq:ops_1}  from the $Z-$pole run of FCCee and CEPC studying a mono$-\gamma$ signal in DY production through $s-$channel $Z$ exchange. We assume an integrated luminosity for CEPC and FCC-ee ${\cal L}=16\,$ab$^{-1}$ and ${\cal L}=120\,$ab$^{-1}$, respectively\footnote{See also Ref.\,\cite{Jin:2017guz} for a previous  related study at CEPC.} 

The analysis proceeds along similar lines to the one for the muon collider described in Sec.\,\ref{sec:muC_DY}. We impose at generation level the requirement 
$$ p_T^\gamma>5 \textrm{ GeV}\,, \quad|\eta_\gamma|<2.5\,,$$
which we assume to be the lower threshold for identifying and reconstructing a photon in the detector acceptance.

We take 
$$e^+ e^- \to \gamma \nu\bar \nu $$
as dominant background 
and then maximimze the reach on the Rayleigh operators by optimizing the statistical significance over a $p_T^\gamma$ selection.
The signal, computed for $m_\phi=1\,$GeV and ${\tilde{{\cal C}}}_{{\cal W}}=0$, and background cross-sections with $p_{T,{\rm min}}^{\rm \gamma}>37.5\,$GeV are respectively $\sigma_{\rm sig}\simeq 0.79\,\left(\frac{{\tilde{{\cal C}}}_{{\cal B}}}{10^{-6}\,{\rm GeV}^{-2}}\right)^2\,{\rm fb}$ and $\sigma_{\rm bkg}=0.34\,$fb.

\begin{figure}[t!]
    \centering
    \includegraphics[width=0.45\linewidth]{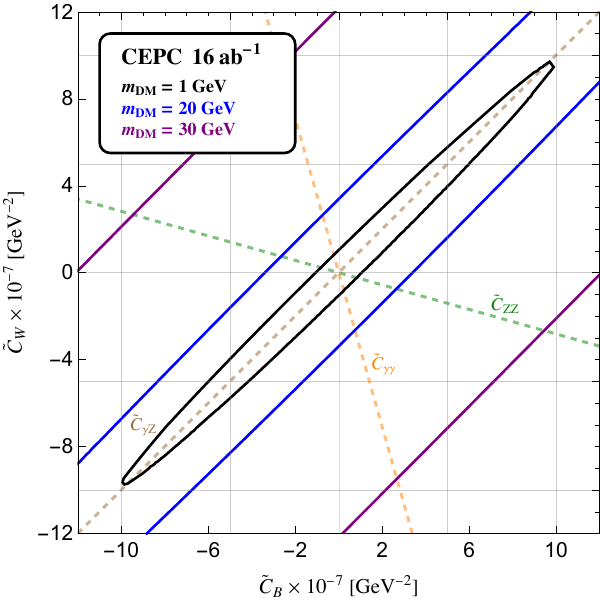}\hspace{0.5cm}
    \includegraphics[width=0.45\linewidth]{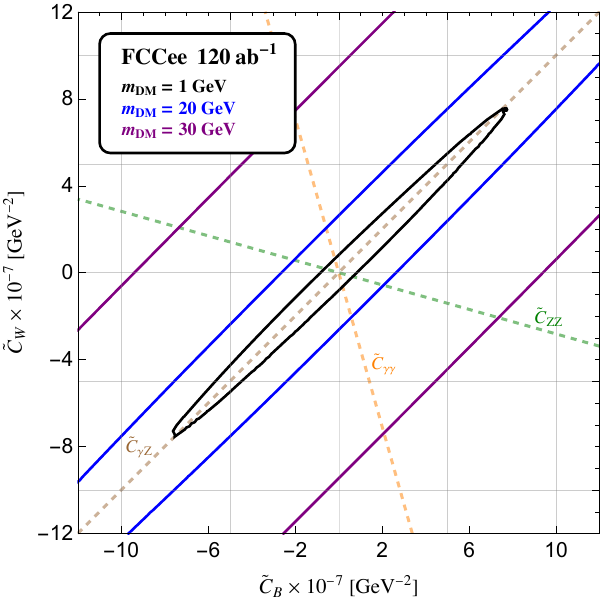} \\
\caption{Constraints in the ${\tilde {\cal C}}_{\cal B}-{\tilde {\cal C}}_{\cal W}$ plane for the DY process at an $e^\pm$ collider operating at the Z-pole, comparing a CEPC-like machine (\textit{left panel}) and an FCCee-like machine (\textit{right panel}), shown for various DM mass values. The dashed lines indicate the directions associated with the couplings defined in Eq.~\ref{eq:EWSB-coeff}. As expected, the ellipses are tilted toward the ${\tilde {\cal C}}_{\gamma \rm Z}=0$ direction, since at the Z-pole, the strongest sensitivity arises from the on-shell Z boson.}

    \label{fig:FCCee-planes}
\end{figure}

The results are reported in Tab.\,\ref{tab:cepc} for 
some representative mass values, assuming ${\cal \tilde C}_{\cal W}=0$ for display. Two integrated luminosities are presented in the table. Similarly, in Fig.\,\ref{fig:FCCee-planes}  we show the expected sensitivity projected in the 
$({\tilde {\cal C}}_{\cal B},{\tilde {\cal C}}_{\cal W})$ plane, as to encompass the most general couplings structure for Rayleigh DM.

In contrast to the case of a hadronic collider, in the case of a lepton collider, the incoming momentum in the effective vertex is well-known and given by \(\sqrt{s}\). Therefore, to ensure the validity of the EFT, in addition to the requirement in Eq.~\eqref{eq:EFTcondition}, we can also \textit{a posteriori} check that \(\sqrt{s} < \Lambda\).

Assuming the UV completion of Eq.~\eqref{eq:ops_1}, we find that the bounds do not comply with the requirement for the validity of the effective description, \(\Lambda_\text{loop} > m_Z\), with \(g_{\rm loop} \sim 4\pi\), for DM masses above 40 GeV for both CEPC and FCCee. On the contrary, the EFT remains valid across the entire mass range for the UV completions relevant for Eq.~\eqref{eq:ops_1_strong}.

In addition we observed a feature of the results in Fig.\,\ref{fig:FCCee-planes}. Along the  ${\tilde {\cal C}}_{\gamma Z}=0$ direction, differently from the LHC case shown in Fig.~\ref{fig:LHC-planes}, there is a significangly weaker sensitivity.
The $Z$ factory sensitivity contours are compressed along the ${\tilde {\cal C}}_{\gamma\gamma}=0$ direction, meaning that for both FCC-ee and CEPC the primary contribution to the signal cross-section originates from the resonantly enhanced $\tilde {{\cal C}}_{\gamma Z}$ operator in Eq.~\eqref{eq:EWSB-coeff}.%, making these collider  effective probes of this interaction. 

\subsection{High energy \texorpdfstring{$\mu$}{TEXT}C}

Interest in a collider using high-energy muons instead of electrons has been growing in recent years. Such machine allows for precise studies of SM processes, benefiting from the clean environment of a lepton machine, and enables direct and efficient exploration of the multi-TeV scale. Muons can be accelerated to higher energies than electrons and, unlike protons, they impress the full center-of-mass energy in the hard scattering, due to the absence of constituents (the quarks and gluons) which end up carrying only a fraction of the center of mass energy. The feasibility of a muon collider is still uncertain due to significant technological challenges. Despite this, the high-energy physics community is actively exploring its potential for SM and Beyond Standard Model (BSM) studies (see reviews such as \cite{AlAli:2021let,MuonCollider:2022xlm,Accettura:2023ked}).

We analyze how the effective interactions of Eq.~\eqref{eq:ops_1} can be tested at a $\mu$C operating at $\sqrt{s}=10$ TeV, which is currently considered as a challenging, yet realistic, center of mass energy. For the target integrated luminosity, we use the commonly adopted luminosity target  $\mathcal{L}=10\,{\rm ab}^{-1}$.

\subsubsection{Drell-Yan  }
\label{sec:muC_DY}
\begin{figure}[t!]
    \centering
        \includegraphics[width=0.46\linewidth]{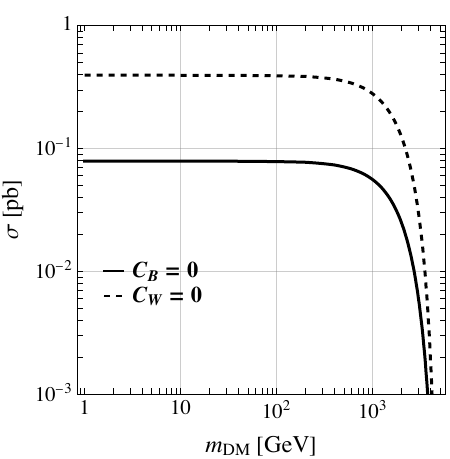}\hspace{0.1cm}
    \includegraphics[width=0.48\linewidth]{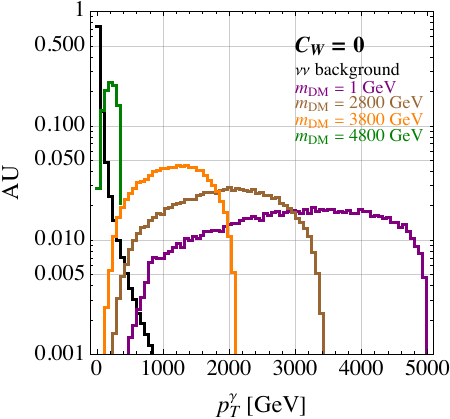} \\
    \caption{
        \textbf{Left Panel:} The DY signal cross-section as a function of the DM mass at the 10~TeV muon collider. The solid and dashed lines for $\tilde{\mathcal{C}}_\mathcal{B}=0$ and  $\tilde{\mathcal{C}}_\mathcal{W}=0$, respectively, with the other coefficient put at $1 \,{\rm TeV}^{-2}$. \textbf{Right panel:} Unit-normalized  \(p_T^\gamma\) distribution for the main background and the signal for several illustrative values of the DM mass for \(\Tilde{C}_\mathcal{W}=0\).
    }
    \label{fig:xsecs_pt_DY}
\end{figure}

At a $\mu$C DM pairs can be produced  with the same mechanism of Fig.\,\ref{fig:LHC_DY} from initial state   muons in place of quarks.  We perform a simple mono$-\gamma$ analysis, for which the dominant SM background, analogoulsy to the hadron collider situation, is  represented by the process $$\mu^+\mu^-\to \gamma\nu\bar\nu \,.$$  
We generated LO signal and background events with generator level cuts $|\eta^\gamma|<2.5$  as to account for the angular acceptance of the proposed detectors at the muon collider. The photon transverse momentum is selected as to maximize the reach for new physics, as described below.
With these requirements the SM background cross-section is estimated to be 4.2\,pb. In Fig.~\ref{fig:xsecs_pt_DY}, we show the signal cross-sections as a function of the DM mass obtained by switching on one Wilson coefficient at the time. For Wilson coefficients of order $1 \,{\rm TeV}^{-2}$ we get cross-section of order 0.1~pb. Neglecting selection on the phase-space effects, the signal cross-section scales as $s^2 {\cal C}_{{\cal B},{\cal W}}^2$.

As for the case of the FCC-hh search of Sec.\,\ref{sec:FCC-hh}, we optimize the reach of the mono$-\gamma$ analysis by only working on the $p_T^\gamma$ variable. We show in Fig.\,\ref{fig:xsecs_pt_DY} the $p_T^\gamma$ distribution for the main SM background and for the new physics signal for some representative DM mass values.

In contrast to the hadron collider cases (see Fig.\,\ref{fig:xsecs_pt_DY}), we observe that the signal $p_T^\gamma$ distribution shifts to lower values as the DM mass increases. This behavior is a consequence of energy conservation and the fact that, at the muon collider, the entirety of the center-of-mass energy participates in the hard-scattering process. As a result, the analysis cut is optimized at lower $p_T^\gamma$ values for higher DM masses. 

We   report in Tab.\,\ref{tab:muC-optimal_cut_DY} the optimal cut on $p_T^\gamma$ and the sensitivity we obtain on  ${\cal \tilde C}^{-1/2}_{{\cal B}}$ as defined in Eq.\,\eqref{eq:strong_matching} obtained by fixing for ${\tilde {\cal C}}_{{\cal W}}=0$ for various DM masses.
As previously noticed, differently from the hadron-collider cases, here to comply with the validity of the effective description for the operators of Eq.\,\eqref{eq:ops_1_strong}
is enough to check a posteriori that $\Lambda_{\rm tree}> \sqrt s$. Choosing $g_{\rm tree}\sim 4\pi$ we can find that the EFT approach breaks for DM masses around 4 TeV.

\begin{table}[t!]
\centering
\rowcolors{1}{gray!10!white}{white} % Alterna i colori delle righe
\begin{tabular}{ c @{\hskip 10pt} | @{\hskip 10pt} c @{\hskip 10pt} | @{\hskip 10pt} c }
\toprule
\rowcolor{white!10} 
\multicolumn{3}{c}{mono-$\gamma$ DY,~~$\sqrt{s}=10\,$TeV ~~${\tilde{{\cal C}}}_{\cal W}=0$}~~$\mu$C~~$({\cal L}=10\,$ab$^{-1})$ \\
\midrule
%\rowcolor{red!30!orange!50}
\rowcolor{gray!20}
$m_{\rm DM}\,$[GeV] & $p^\gamma_{T,\rm min}\,$[GeV] & ${\tilde {\cal C}}^{-1/2}_{\cal B}\,$[GeV] \\
\midrule
\rowcolor{white!10} 
 200 & 3000 & 10541 \\
\rowcolor{white!20}  
 1000 & 2800 & 9440 \\
 \rowcolor{white!20}  
 2000 & 2300 & 7177 \\
 \rowcolor{red!20}  
 4000 & 900 & 1841\\
 \bottomrule
\end{tabular}
\caption{
Bounds on \(\Lambda_{\rm tree} = {\tilde {\cal C}}^{-1/2}_{\cal B}\) with \({\tilde {\cal C}}_{\cal W}=0\) from Drell-Yan processes at a muon collider with a center of mass energy \(\sqrt{s} = 10\) TeV for various DM masses are presented. It is straightforward to observe that the validity of the effective field theory for the mono-photon search at the muon lepton collider is violated as soon as \(\Lambda_{\rm tree}\) approach \(\sqrt{s} = 10\) TeV. We have highlighted in red the mass for which the EFT breaks down, for a given choice of the coupling $g_{\rm tree}\sim 4\pi$. 
}
\label{tab:muC-optimal_cut_DY}
\end{table}

\subsubsection{Vector boson scattering}

\begin{figure}[t!]
    \centering
\includegraphics[width=0.99\linewidth]{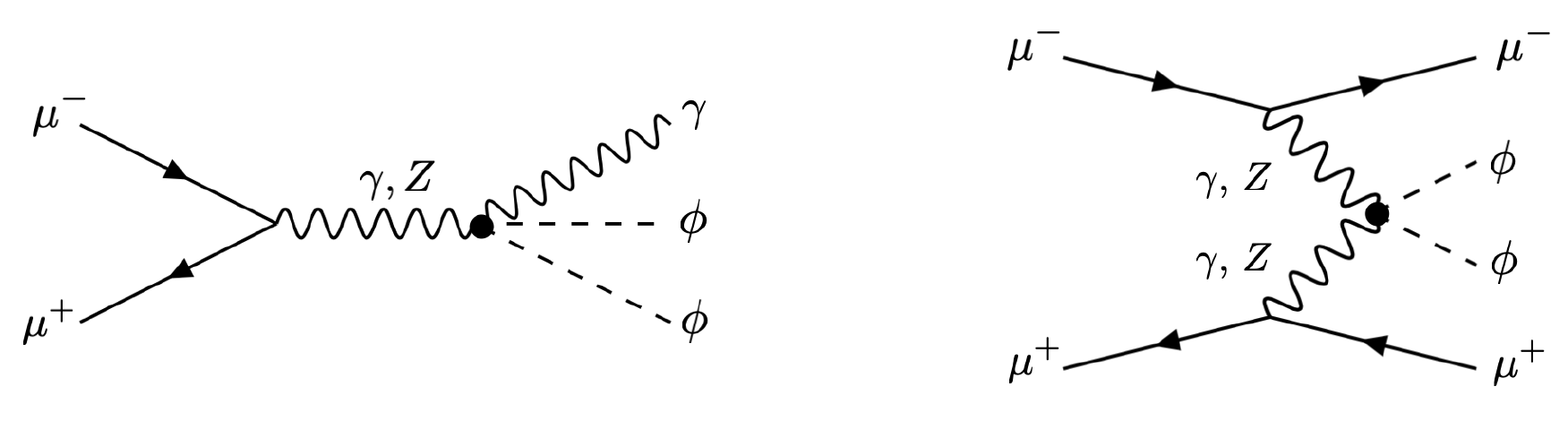}  
\caption{
Representative Feynman diagrams for 
DY-like (left) and 
neutral VBS-like (right)  pair-production of scalar $\phi$ particles through the operator of Eq.\,\eqref{eq:ops_1}. 
}
\label{fig:muC_VBS}
\end{figure}

An interesting characteristic of  a $\mu$C
operating at multi-TeV energies is that it will be 
able to  effectively scatter EW gauge bosons radiated from the colliding $\mu$ beams\,\cite{Costantini:2020stv,Han:2020uid,Ruiz:2021tdt,Garosi:2023bvq}, thus effectively acting as a gauge boson collider.
If the scattered gauge bosons are the SM $\gamma$ and/or $Z$, 
the remnant of the splitting are two muons, that can in principle be detected in order to tag this type of scattering.
This can offer an alternative probe to test the 
effective interactions of Eq.\,\eqref{eq:ops_1}
in neutral vector boson scattering processes as 
\begin{equation}
    \mu^+ \mu^-\to \mu^+ \mu^- \phi\phi\,, \label{eq:VBSmuCol}
\end{equation}
 whose  Feynman diagrams are shown in  Fig.\,\ref{fig:muC_VBS}.
We note that the signal gets contributions from both VBS-like diagrams (left) and DY-like (right) diagrams with a final state gauge boson  splitting into a muon pair. We have checked that the VBS-like production dominates over the DY-like one when the angular coverage of the detector is  extended to small polar angles, as we will do in this section\footnote{See {\it{e.g.}} Ref.~\,\cite{Acanfora:2023gzr} for a similar discussion for the case of the Belle~II experiment and possible extensions of this detector in the forward direction, which may precede (hence prove the concept of) an experiment of this type at a future muon collider.}. 
 Moreover, we find that as the DM mass increases the DY cross-section decreases more rapidly than the VBS one, at fixed angular coverage of the detector. This is  to be expected because of the derivative nature of the effective Rayleigh coupling and its growing-with-energy behavior. In fact, the energy entering the effective vertex differs between the DY and the VBS processes. In the former case the momentum exchanged is always fixed to $\sqrt{s}$=10~TeV, while in the latter case
 it is equal to the missing invariant mass (MIM), which satisfies ${\rm MIM} > 2 m_{\rm DM}$.

The kinematic of Eq.~\eqref{eq:VBSmuCol} features two forward muons, with a small polar angle with respect to the beam direction, flying in opposite directions. 
In order to be sensitive to this  process the detector need to be sensitive to  $\mu^\pm$  produced with a high absolute pseudo-rapidity. The possibility of detecting forward $\mu^\pm$ with $|\eta|$ up to  7
have been  discussed in\,\cite{Ruhdorfer:2023uea, Ruhdorfer:2024dgz} in the context of invisibly decaying SM Higgs bosons produced in neutral VBS. The authors propose to install a dedicated forward muon spectrometer to tag events where the $\mu^\pm$ beams radiate a neutral EW gauge boson that participates in the hard scattering.

To assess the sensitivity to VBS production of DM, we have simulated both the signal Eq.~\eqref{eq:VBSmuCol} and the SM background 
\begin{equation}
     \mu^+\mu^- \to \mu^+\mu^- \nu \bar \nu\,,
    \label{eq:background-mucol} 
    \end{equation}
at LO with 
{\tt MadGraph5\_aMC@NLO} for $\sqrt s=10\,$TeV.

We remark that for our exploratory goal we pursue a simplified background treatment, without inclusion of detector resolution effects, thus less realistic than  what has been done in recent works \cite{Ruhdorfer:2023uea, Ruhdorfer:2024dgz}. 
These works cover a region of phase-space at lower MIM, however, they can be used for an estimate of the size of the background in the region of phase-space relevant for our analysis.For MIM below about 1~TeV, Ref.~\cite{Ruhdorfer:2024dgz} finds that as the MIM increases the background composition becomes progressively dominated by the process identified in Eq.~\eqref{eq:background-mucol}.
Other background sources include $\mu\mu\gamma$, $\mu\mu f\bar{f}$, and $\mu\mu WW$ final states, 
where the $\gamma$, $f\bar{f}$, or $WW$ remain unobserved. Consequently, the $\mu\mu\nu\nu$ 
background contributes approximately $1/6$ of the total rate for MIM around 1~TeV. Starting from this observation we extrapolate towards phase-space characterized by large MIMs, where our signal is more evident, and the fraction of $\mu\mu \nu \nu$, based on the results of Ref.~\cite{Ruhdorfer:2024dgz}, is expected to be larger than $1/6$. To approximately account for all sub-leading background processes in our $\rm MIM>3\,TeV$ analysis we estimate the total background to be 6 times the LO cross-section for $\mu\mu \nu \nu$. This should more than abundantly account for other sources of background. While this magnification of the background may be overly exaggerated, we find that the reach of the muon collider is in any case very competitive. Therefore, in Fig.~\ref{fig:fig_sc_finalplot} and Fig.~\ref{fig:fig_wc_finalplot} we will present our result in this set-up, and leave a more refined determination of the background at high MIM including detector effects to future work.

To perform our analysis, following Ref.\,\cite{Ruhdorfer:2023uea, Ruhdorfer:2024dgz} about the characteristics of the forward detector, we impose at
generation level an energy threshold for the muons to be detected, 
\begin{equation}
E_{\mu^\pm}>500~{\rm GeV}. \label{eq:VBSmuColacc-one}
\end{equation}
This requirement is needed in order for the muons to penetrate the conical absorber which is placed along the beam line to shield the detector from the radiation induced by the decay of the colliding leptons. Still motivated 
by Ref.\,\cite{Ruhdorfer:2023uea} we restrict the muon pseudo-rapidity ($\eta$) and their angular separation ($\Delta R$) to have 
\begin{equation}
|\eta_{\mu^\pm}|<7\,,\quad \Delta R(\mu^+\mu^-)>0.4\,. \label{eq:VBSmuColacc-two}
\end{equation} 
%and further require an angular separation between the two muons. 
Additionally, in order to suppress Bhabha scattering, we select only events that satisfy
\begin{equation}
    p_T^{\mu\mu}>50\textrm{ GeV}\,, \label{eq:VBSmuColacc-three}
\end{equation}
 where $p_T^{\mu\mu}$ is the transverse momentum of the di-muon system. 
With these generation level cuts 
 the background rate is $\sigma_{\rm bkg} = 0.157\,$~pb. For comparison, the signal cross-section with ${\cal \tilde C}_{\cal W} = 0$ and $m_{\rm DM}=100\,$GeV is
$\sigma_{\rm sig} =\left({{\tilde{\cal C}}_{\cal B}}/{{\rm TeV}}^{-2}\right)^2$0.645~pb.

\begin{figure}[t!]
    \centering
    \includegraphics[width=0.45\linewidth]{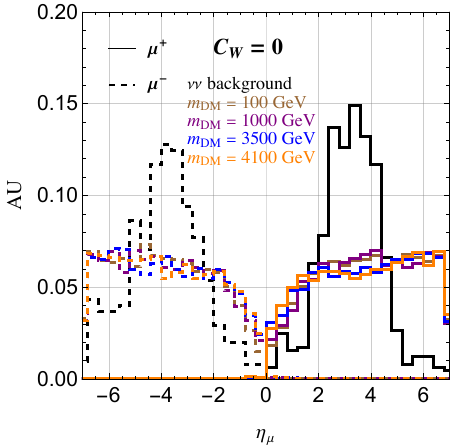}\hspace{0.1cm}
    \includegraphics[width=0.45\linewidth]{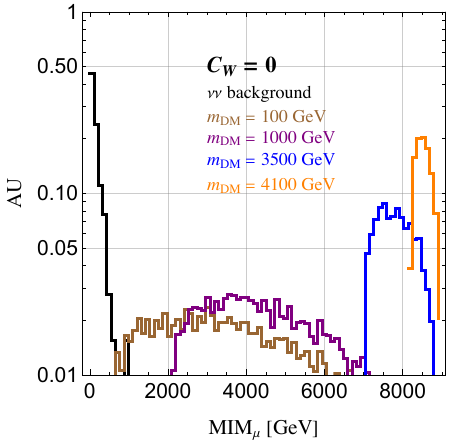}  
    \caption{Normalized distributions for pseudorapidity (\textit{left panel}) and MIM (\textit{right panel}), are shown for a muon collider operating at $\sqrt{s} = 10\,$TeV. The signal distributions, corresponding to different DM masses, are compared with the neutrino background process $\mu^+ \mu^- \to \mu^+ \mu^- \nu \bar{\nu}$, with $\tilde{{\cal C}}_{{\cal W}}=0$.
    \label{fig:distr_muC_baseline}}
\end{figure}

To better characterize the differences between our signal and the SM background we   show in Fig.\,\ref{fig:distr_muC_baseline} the unit-normalized distributions of the the  muons pseudo-rapidity (left panel), considering both charges, and the MIM distribution (right panel), where $
    {\rm MIM} = \sqrt{\Delta p_\mu \Delta p^\mu} $
    with $\Delta p^\mu = (\sqrt s, \vec 0) - p_{\mu^+} - p_{\mu^-}\,$, where 
$p_{\mu^\pm}$ are the four-momenta of the final state muons.

From the figures it is evident that the signal features high-rapidity muons traveling in opposite directions, a property that we use for background rejection.
Inspecting the MIM distributions we see that the background peaks at smaller values with respect to the signal, whose distributions  start at the threshold value MIM$>2m_{\rm DM}$. 
Therefore we require 
\begin{equation}
    {\rm sign}(\eta_{\mu^+}\eta_{\mu^-})<0\,,
\end{equation}
and then compute the reach on the Rayleigh Wilson coefficient by optimizing the selection on the minimum MIM for each DM mass.
An example of the evolution of the rate for signal and background is reported in Tab.~\ref{tab:muC_VBS_cutflow} for $m_{\rm DM}=100\,{\rm GeV}$.

\begin{table}[t!]
    \centering
    \rowcolors{1}{gray!10!white}{white} % Alterna i colori delle righe
    \begin{tabular}{ c @{\hskip 10pt} | @{\hskip 10pt} c  c  c  }
    \toprule
    \rowcolor{white!10} 
    \multicolumn{4}{c}{VBS at $\mu$C~~$\sqrt s=10\,$TeV} \\
    \midrule
    %\rowcolor{red!30!orange!50}
    \rowcolor{gray!20}
    cut flow & $\sigma_{\rm bkg}^{\nu\nu}$ [pb] &$\sigma_{\rm sign}^{{\tilde {\cal C}}_{\cal W}=0}$ [pb] & $\sigma_{\rm sign}^{{\tilde {\cal C}}_{\cal B}=0}$ [pb] \\
    \hline
    \rowcolor{white!10}
    Baseline cuts  Eq.~\eqref{eq:VBSmuColacc-one}-\eqref{eq:VBSmuColacc-three}   &   0.157        &   0.645    &   0.131 \\
    \rowcolor{white!10}
    sign$(\eta_{\mu^+}\eta_{\mu^-})<0$    &    0.130       &   0.621    &  0.127 \\
    \rowcolor{white!10}
    ${\rm MIM}>3$ TeV & 0.0005 & 0.361 & 0.067 \\
     \bottomrule
    \end{tabular}
    \caption{A representative cut-flow for the vector boson scattering analysis at a muon collider with a center of mass energy of \(\sqrt{s} = 10\) TeV and a DM mass of \(m_{\rm DM} = 100\) GeV is presented setting ${\tilde {\cal C}}_{\cal W,B}=0$ and the other Wilson coefficient set to $1 \,$TeV$^{-2}$. 
    This cut flow refers only to the neutrino background channel.
    }
    \label{tab:muC_VBS_cutflow}
    \end{table}

The sensitivity to VBS production of DM is shown in Tab.~\ref{tab:muC-optimal_cut}.
To ensure the validity of the EFT we check a posteriori that MIM does not exceed the scale $\Lambda$ relevant for the UV completion hypothesis. 
As for the LHC case, with the UV completion hypothesis of Eq.~\eqref{eq:ops_1}, one finds that the obtained bounds do not comply with the requirement of the validity of the effective description ${\rm MIM_{\rm min}}<\Lambda_{\rm loop}$, while this can be retained if the operators arise from a UV completion that generates the effective interaction at a scale $\Lambda_{\rm tree}$ from tree-level interactions given by Eq.~\eqref{eq:strong_matching}. 
In this case we further impose the requirement ${\rm MIM}_{\rm min}<\Lambda_{\rm tree}$ in setting our limits.

\begin{table}[t!]
    \centering
    \begin{tabular}{ c @{\hskip 10pt} | @{\hskip 10pt} c c }
    \toprule
    \multicolumn{3}{c}{VBS at $\mu$C~~$\sqrt{s}=10\,$TeV~~${\cal L}=10\,$ab$^{-1}$~~${\tilde{{\cal C}}}_{\cal W}=0$}\\
    \hline
    \rowcolor{gray!10}
    $m_{\rm DM}\,$[GeV] 
    & $\rm MIM_{\rm min}\,$[GeV] 
    & ${\tilde {\cal C}}^{-1/2}_{B}\,$[GeV]\\ 
    \hline
    100 & 3000 & 12769  \\
    1000 & 3500  & 12454  \\
    2000 & 4500  & 11183  \\
    3000 & 6000  & 8607  \\
    \rowcolor{red!20}  
    4000 & 7400  &4558  \\
    \rowcolor{red!20}  
    4400 & 8750  &2461  \\
    \bottomrule
    \end{tabular}
    \caption{
    Bounds on \({\tilde {\cal C}}^{-1/2}_{\cal B}\) with \({\tilde {\cal C}}_{\cal W}=0\) from VBS processes at a muon collider with a center of mass energy of \(\sqrt{s}=10\,\text{TeV}\) for various DM masses are presented. Bounds and optimizations refers only to the neutrino background channel. The effective field theory description becomes invalid when \({\rm MIM_{\rm min}} > \Lambda_{\rm tree}\), and we have highlighted in red the corresponding rows where this condition is met. 
    }
    \label{tab:muC-optimal_cut}
    \end{table}

In Fig.~\ref{fig:fig_sc_finalplot} we show the most stringent bound between DY and VBS processes, taking into account the EFT validity.

\section{Thermal freeze-out}\label{sec:relic}
In this section we focus on thermal DM produced via the so called \textit{thermal freeze-out}. 
The relevant DM annihilations in the early universe are into EW gauge bosons $\phi\phi \leftrightarrow \gamma \gamma,\gamma Z, ZZ, W^+W^-$.
In the non relativistic limit the corresponding velocity averaged thermal cross sections are $s$-wave and given by:
\begin{equation}\label{eq:ann_xsecs}
    \begin{split}
        &\left<\sigma v\right>_{\gamma\gamma}= \frac{8 m^2_{\rm DM}}{\pi}\mathcal{\tilde C}_{\gamma\gamma}^2\, ,\\
        &\left<\sigma v\right>_{\gamma Z}= \frac{4 m^2_{\rm DM}}{\pi}\mathcal{\tilde C}_{\gamma Z}^2\left(1-\frac{m_Z^2}{4 m_{\rm DM}^2}\right)^3\theta(2 m_{\rm DM}-m_Z)\, ,\\
        &\left<\sigma v\right>_{Z Z}=\frac{8 m^2_{\rm DM}}{\pi}\mathcal{\tilde C}_{ZZ}^2\left(1-\frac{m_Z^2}{m_{\rm DM}^2}+\frac{3}{8}\frac{m_{\rm Z}^4}{m_{\rm DM}^4}\right)\sqrt{1-\frac{m_Z^2}{m_{\rm DM}^2}}\theta(m_{\rm DM}-m_Z)\, ,\\
        &\left<\sigma v\right>_{W W}=\frac{4 m^2_{\rm DM}}{\pi}\mathcal{\tilde C}_{WW}^2\left(1-\frac{m_W^2}{m_{\rm DM}^2}+\frac{3}{8}\frac{m_{\rm W}^4}{m_{\rm DM}^4}\right)\sqrt{1-\frac{m_W^2}{m_{\rm DM}^2}}\theta(m_{\rm DM}-m_W)\, .
        \end{split} 
\end{equation}

Following the approach outlined in Ref.~\cite{Kolb:1990vq}, we solve the Boltzmann equation for thermal freeze-out, including all channels in Eq.~\eqref{eq:ann_xsecs} that are kinematically accessible. By requiring the relic density to satisfy $\Omega h^2 \simeq 0.12$~\cite{Planck:2018vyg}, we determine lines in the ($m_{\rm DM}$,\,${\tilde {\cal C}_{{\cal B,W}}}$) planes. These results are visualized in several figures by selecting ${\tilde {\cal C}}_{\cal B}=0$ \textit{left panels} and ${\tilde {\cal C}}_{\cal W}=0$ \textit{right panels}. 
For instance, in Fig.~\ref{fig:fig_sc_finalplot} we display the thermal line benchmark with all the current and future experimental probes of Rayleigh DM.

 As the DM mass increases, additional channels involving massive final state gauge bosons become accessible, leading to a larger annihilation cross-section for fixed couplings. 
A notable feature of the relic abundance curves, particularly evident for ${\tilde {\cal C}}_{\cal B}=0$, is the emergence of bumps at the mass thresholds corresponding to the processes in Eq.\,\eqref{eq:ann_xsecs}. As the cross section becomes larger we need larger scale to maintain the relic value fixed.

\section{Direct detection bounds}\label{sec:DD}
In this Section, we summarize the relevant bounds derived from DM direct detection experiments based on nuclear recoil processes. As previously mentioned, Rayleigh DM is particularly elusive. This is primarily due to the fact that it does not couple to the relevant degrees of freedom in DD such as light quarks and gluons, resulting in DM-nucleon cross section that is loop-suppressed. However, the running and matching of the operators $\phi^2 B_{\mu\nu}B^{\mu\nu}$ and $\phi^2 W^a_{\mu\nu}W^{a,\mu\nu}$ with those defined below the EW scale, induced extra couplings with light degrees of freedom.
To facilitate our analysis, it is useful to define an effective field theory at the nuclear scale energy $m_{\rm N}$ defined as:
\begin{equation}
    \mathcal{L} \supset \sum_{q=u,d,s} {\tilde {\cal C}}_q m_q \phi^2 \bar{q} q + {\tilde {\cal C}}_G \phi^2 G^{A,\mu\nu} G_{\mu\nu}^A + {\tilde {\cal C}}_{\gamma\gamma} \phi^2 F^{\mu\nu} F_{\mu\nu}\, ,
    \label{eq:DDeq}
\end{equation}
where $G^{A,\mu\nu}$ is the gluon field strength tensor, the sum runs over the light quarks $q=u,d,s$ with mass $m_q$ and the Wilson coefficient.
By performing the renormalization group evolution, we match the coefficients $\tilde{{\cal C}}_{\cal B}$ and $\tilde{{\cal C}}_{\cal W}$ defined at the EW scale down to the nuclear scale. This process generates new couplings as follows:
\[
({\tilde {\cal C}}_{\mathcal W},{\tilde {\cal C}}_{\mathcal B})_{\mu=v}~~\xrightarrow{\text{RGE}}~~ ({\tilde {\cal C}}_q,{\tilde {\cal C}}_G,{\tilde {\cal C}}_{\gamma\gamma})_{\mu=m_{\rm N}}.
\]
Following the procedure outlined in \cite{PanciDeramo}, the coefficients at the nuclear scale are defined as:
\begin{equation}
    \begin{split}
        {\tilde{\cal C}}_u(m_{\rm N}) &\simeq -0.046\,{\tilde{\cal C}}_{{\cal B}}(\Lambda) + 0.15\,{\tilde{\cal C}}_{{\cal W}}(\Lambda),\\
        {\tilde{\cal C}}_{d,s}(m_{\rm N}) &\simeq -0.021\,{\tilde{\cal C}}_{{\cal B}}(\Lambda) + 0.14\,{\tilde{\cal C}}_{{\cal W}}(\Lambda),\\
        {\tilde{\cal C}}_{GG}(m_{\rm N}) &\simeq 5.5 \times 10^{-4}\,{\tilde{\cal C}}_{{\cal B}}(\Lambda) + 2.5 \times 10^{-3}\,{\tilde{\cal C}}_{{\cal W}}(\Lambda),\\
        {\tilde{\cal C}}_{\gamma\gamma}(m_{\rm N}) &\simeq 0.77\,{\tilde{\cal C}}_{{\cal B}}(\Lambda) + 0.23\,{\tilde{\cal C}}_{{\cal W}}(\Lambda).
    \end{split}
\end{equation}
Having at our disposal the interaction lagrangian, we compute the matrix element at the nucleon level by dressing Eq.~\eqref{eq:DDeq} with the initial and final states involved in the scattering process. Furthermore, since the nucleus is not a point-like object, it is necessary to correct the DM-nucleon matrix element using nuclear form factors: \textit{i)} the Helm form factor, $F_{\rm H}$, represents the response initially originated by the dressing of quark and gluon currents, while the \textit{ii)} Rayleigh response, $F_{\rm ray}$, originating from the photon coupling. Both form factors are normalized to one in the limit $q \to 0$ and depend on the exchanged momentum $q$ through the dimensionless variable $y = (qb/2)^2$. Here, the parameter $b$ is given by $b\approx1.3\,(1.8A_{\rm T}^{1/3}-1)^{-1/2}\,\rm{fm}$ where $A_{\rm T}$ denotes the mass number of the target nucleus.  
Explicit expressions for these form factors can be found in~\cite{LiamFitzpatrick_2013} and~\cite{Frandsen_2012}.

We checked that the main contribution to the Rayleigh cross section primarily arises from the coupling with photons and therefore we neglect the contribution from quark and gluon couplings. The differential cross section for Rayleigh DM in the deeply non-relativistic regime simplifies to:  
\begin{equation}
    \frac{d\sigma^{\text{Ray}}}{dq^2} \simeq \frac{2 \alpha_{\rm em}^2}{\pi^2 b^2 m_{\rm DM}^2 v_{\rm DM}^2} \tilde{{\cal C}}_{\gamma\gamma}^2 Z_{\rm T}^4 F^2_{\text{ray}}(y),
    \label{eq:diffRayxsec}
\end{equation}
where \(m_{\rm T}\) and \(Z_{\text{T}}\) represent the target mass and atomic number, respectively, \(\alpha_{\text{em}}=e^2/4\pi\) is the electromagnetic fine-structure constant, and \(v_{\text{DM}}\) denotes the DM-target nucleus relative velocity. 

As one can see, the dependence on recoil energy is encapsulated entirely within the form factor, analogously to the SI interaction. Indeed it  
reads
\begin{equation}
    \frac{d\sigma^{\rm SI}}{dq^2} = \frac{1}{4 \mu_{\rm DM,n}^2 v_{\text{DM}}^2} \sigma^{\rm SI}_n A_{\rm T}^2 F^2_\text{H}(y),
    \label{eq:diffSIxsec}
\end{equation}
where $\sigma^{\rm SI}_n$ represents the DM-nucleon SI cross section and \(\mu_{\rm DM,n}\) is the DM-nucleon reduced mass and $\mu_{\rm DM,n}\sim m_{\rm DM}$ for $m_{\rm DM}\ll m_{\rm n}$. Usually, the experimental collaboration provides upper limits on the spin-independent WIMP-nucleon cross section at the \(90\%\) C.L.

In general, Rayleigh and Helm form factors have a different momentum dependence behavior  and therefore it is useful to define a reference momentum $\bar{q} = 2 \mu_{\rm DM,\text{T}} v_0$, and therefore $\bar{y}$ to easily derive the bound. Here, $\mu_{\rm DM,\text{T}}$ is the DM-target reduced mass  and $v_0 = 220\,\si{km\,s^{-1}}$ represents the mean local DM velocity. Comparing Eq.~\eqref{eq:diffRayxsec} and Eq.~\eqref{eq:diffSIxsec} an approximate bound reads
\begin{equation}
    {\tilde {\cal C}_{\gamma\gamma}}\simeq \frac{\pi}{2\sqrt{2}}\frac{A_{\rm T}}{Z_{\rm T}^2}\frac{|F_{\rm H}(\bar y)|}{|F_{\rm ray}(\bar y)|}\sqrt{\sigma^{\rm SI}_n} \ .
    \label{eq:cgammaDD}
\end{equation}

It is important to notice that in the limit of $m_{\rm DM}\ll m_{\rm T}$ the $y$ parameter tends to zero and both form factors tend to 1. In this case there is no $q$-dependence in~\eqref{eq:cgammaDD} leading to an exact rescaling of the experimental bound.

In this work we used experimental limits on the SI WIMP-nucleon cross section provided by the LUX-ZEPLIN (LZ) experiment. This experiment employs a liquid xenon (LXe) time projection chamber (TPC) with a fiducial mass of 5.5 tons. The reported data correspond to a 60-day exposure period and provide the upper limit on the SI WIMP-nucleon cross section at the 90\% confidence level, as detailed in \cite{PhysRevLett.131.041002}.

For the projected sensitivity related to next-generation detectors, we consider the proposed XLZD experiment led by the LZ, XENON, and DARWIN collaborations. The experiment is expected to achieve an exposure of 200 tonne-years, potentially reaching the sensitivity limit imposed by astrophysical neutrinos~\cite{XLZD:2024nsu}, often referred to as the “neutrino fog”~\cite{OHare:2021utq}. These projected limits are used to explore future constraints on DM interactions. 

Our results are depicted in Fig.~\ref{fig:fig_sc_finalplot} and Fig.~\ref{fig:DDandIDbound} with solid (LZ-2022) and dashed (XLZD) gray lines. In the left panel we show the results for ${\tilde {\cal C}}_{\cal B}=0$ and in the right panel for ${\tilde {\cal C}}_{\cal W}=0$. As one can notice, since the contribution arises only from the ${\tilde {\cal C}}_{\gamma\gamma}$ the results in the two cases  just differ by the tangent of the Weinberg angle as one can infer from~\eqref{eq:EWSB-coeff}. The current sensitivity is weaker with respect to the current LHC limits, while in the case of ${\tilde {\cal C}}_{\cal W}=0$ for DM masses around $\sim 10\, \rm GeV$ it is slightly stronger and comparable with HL-LHC projections. Regarding future DD probes, the sensitivity will improve substantially reaching in the ${\tilde {\cal C}}_{\cal W}=0$ case the current ID limit.

\section{Indirect Detection}\label{sec:astro}
In this section, we examine indirect detection probes resulting from DM annihilation of Rayleigh DM. Particularly relevant are the annihilation channels into $\gamma\gamma$ and $\gamma Z$, which produce sharp spectral features that maximize the experimental sensitivity. In our analysis, we compute the limits on $\gamma$-ray lines imposed by 15 years of FERMI-LAT~\cite{Fermi-LAT:2009ihh} data as well as the expected sensitivity of the upcoming Cherenkov Telescope Array (CTA)~\cite{CTAConsortium:2010umy}.

\subsection{FERMI-LAT Constraints}
\subsection*{Dataset Specifications and Extraction}\label{sec:astro_dataset}

In this work, we analyze approximately 180 months of data from August 4, 2008, to July 20, 2023, selecting CLEAN events from the PASS8 dataset. We include only high-quality time intervals removing periods when the LAT operated at rocking angles $\theta_r> 52$ deg. The Earth's limb contamination is reduced applying a zenith-angle cut of $\theta_z < 90^\circ$. Our analysis focuses on EDISP3 events (\texttt{evtype = 512}) and \texttt{P8R3 CLEAN V3} version of the instrument response functions. Data extraction and exposure map calculations are performed using the latest version of the \texttt{ScienceTools} software~\cite{FermiTools}. 

Within these specifications, we extract data spanning from 0.5 GeV to 480 GeV in 210 logarithmically spaced bins. Furthermore, following the previous analyses of the FERMI collaboration~\cite{Fermi-LAT:2013thd}, we consider specific regions of interests (RoIs) of the inner Galaxy, which have been found to optimize the signal to noise ratio for line searches. In this paper we focus on RoI41, i.e. a circular region with angular aperture of $41^\circ$ around the Galactic Center.  To this respect, we subtract the regions of 68\% containment around known point sources by using the $4FGL\_DR2$ catalog and we mask the galactic plane in the $|b|<5^\circ,\,|\ell|>6^\circ$ region, with $(b,\ell)$ denoting the set of galactic coordinates.

Further details about the data extraction can be found in Ref.~\cite{DeLaTorreLuque:2023fyg}.

\subsection*{Data Analysis}
In order to extract the constraints, we first compute the photon flux arising from DM annihilation. Since we are interested in the constraints on $\gamma$-ray lines, the relevant annihilation channels are in $\gamma\gamma$ and $\gamma Z$ final states. Both of these channels contribute to the production of spectral lines due to the photons in the final state. In particular, the $\gamma \gamma$ channel produces a spectral line peaked at $E_{\gamma\gamma}\approx m_{\rm DM}$, while the $\gamma Z$ channel produces a second spectral line peaked at lower energy, $E_{\gamma Z}\approx m_{\rm DM}(1-m_Z^2/4m_{\rm DM}^2)$. At the practical level, the differential $\gamma$-ray flux $d\Phi/dE_{\gamma}$ at Earth of self-conjugated DM particles reads
\begin{equation}\label{eq:flux}
    \frac{d\Phi(E_{\gamma})}{dE_{\gamma}}=\frac{\mathcal{J}(\Delta\Omega)}{8\pi m_{\rm DM}^2}\left[\left<\sigma v\right>_{\gamma\gamma}\frac{dN^{\gamma}_{\gamma}}{dE_{\gamma}}(E_{\gamma\gamma})+\frac{1}{2}\left<\sigma v\right>_{\gamma Z}\frac{dN^{\gamma}_{\gamma}}{dE_{\gamma}}(E_{\gamma Z})\right],
    \end{equation}
    where the thermally averaged annihilation cross-sections $\left<\sigma v\right>_{\gamma\gamma,\gamma Z}$ are given in Eq.~\eqref{eq:ann_xsecs}, while $dN^{\gamma}_{\gamma}/dE_{\gamma}(E_{\gamma\gamma,\gamma Z})$ denotes the photon energy spectrum per one annihilation in the channel containing photons in the final state with energy $E_{\gamma\gamma,\gamma Z}$, computed using the tools from Ref.~\cite{Cirelli:2010xx}. Finally, $\mathcal{J}(\Delta \Omega)$ is the annihilation $J$-factor  (i.e., the integral along the line of sight of $\rho^2(r)$, with $\rho(r)$ being the DM density profile) integrated in a region with angular aperture $\Delta \Omega$. The $J$-factor depends on the specific RoI and the assumed DM density profile, and we choose RoI41 of the inner Galaxy, as outlined in Sec.~\ref{sec:astro_dataset}. For this region the FERMI-LAT collaboration found that the DM profile that optimizes the signal-to-noise  ratio is an NFW profile\footnote{The scale radius is $r_s= 20$ kpc, while the scale density $\rho_s\simeq 0.35\,\text{GeV}\text{cm}^{-3}$ is fixed by demanding that the local DM density at the solar radius is $\rho(r_{\odot}=8.5\,\text{kpc})=0.4\,\text{GeV}\text{cm}^{-3}$.}, which gives a $J$-factor  $\mathcal{J}^{\rm NFW}_{41}= 8.53 \times 10^{22}\,\text{GeV}^2\text{cm}^{-5}$~\cite{Fermi-LAT:2013thd}.

We now derive the constraints from the double-line signals described above. Following prior analyses in the literature~\cite{Fermi-LAT:2013thd, DeLaTorreLuque:2023fyg, Foster:2022nva}, we examine a DM mass range from 10 GeV to 300 GeV and apply the \textit{sliding-energy window technique}~\cite{Fermi-LAT:2013thd, Weniger:2012tx}. Specifically, for each DM mass, we perform a likelihood analysis over an energy window centered around $E_{\gamma\gamma} = m_{\rm DM}$ with a width defined by (see e.g. Ref.~\cite{DeLaTorreLuque:2023fyg}) 
\begin{equation}
    \operatorname{min} \left[E_{\gamma Z},\, E_{\gamma\gamma} - 3\sigma(E_{\gamma\gamma}) E_{\gamma\gamma} \right] \le E_{\gamma} \le E_{\gamma\gamma} + 3\sigma(E_{\gamma\gamma}) E_{\gamma\gamma}.
\end{equation}
Here, $\sigma(E)$ represents the half-width of the 68\% acceptance-weighted energy resolution, as specified in~\cite{FermiReso}. Note that the lower edge of the energy window is chosen to capture the double-line features. Additionally, we have also checked that enlarging the window up to $5 \sigma(E)$ affects the overall results only by a few percent. 

Within this energy window, we assume the background model for the number count rate follows a power-law, so that the theoretical background count in the $i$-th energy bin is given by
\begin{equation}
    n^i_{\rm bkg} = A_{\rm bkg} E_i^{\gamma_{\rm bkg}}
\end{equation}
where $A_{\rm bkg}$ and $\gamma_{\rm bkg}$ are two free parameters that control the background amplitude and the spectral index, respectively. Regarding the DM signal count, it is determined by convolving Eq.~\eqref{eq:flux} with the \textit{detector response matrix} (DRM) and the effective exposure $\mathcal{E}$ (defined as the product of effective area and exposure). Both the instrumental response functions are obtained by using the \texttt{FermiTools}. This approach yields the reconstructed number of events in the $i$-th energy bin as follows (see, e.g. Ref.~\cite{Foster:2022nva})

\begin{equation}
	n^i_{\rm DM}=\text{DRM}^{i}_{j}\int_{E_j}^{E_{j+1}}d E_{\gamma} \frac{d\Phi(E_{\gamma})}{dE_{\gamma}}\mathcal{E}(E_{\gamma}) \ .
\end{equation}

Once the number counts have been computed, for each DM mass we build a joint Poissonian likelihood over all the $\mathcal{N}$ energy bins of the specific window

\begin{equation}\label{eq:like_fermi}
	\mathcal{L}({\tilde{{{\cal{C}}}}_{{\cal B}}},{\tilde{{{\cal{C}}}}_{{\cal W}}}, A_{\rm bkg},\gamma_{\rm bkg})=\prod_{i=1}^{\mathcal{N}}\frac{(n_{\text{th}}^{i}({\tilde{{{\cal{C}}}}_{{\cal B}}},{\tilde{{{\cal{C}}}}_{{\cal W}}}, A_{\rm bkg},\gamma_{\rm bkg}))^{n_{\text{obs}}^{i}}}{n_{\text{obs}}^{i}!}e^{-n_{\text{th}}^{i}({\tilde{{{\cal{C}}}}_{{\cal B}}},{\tilde{{{\cal{C}}}}_{{\cal W}}}, A_{\rm bkg},\gamma_{\rm bkg})} \ ,
\end{equation}
where the predicted and observed number count in the $i$-th energy bin are $n_{\rm th}^i\equiv n_{\rm DM}^i+n_{\rm bkg}^i$ and $n_{\rm obs}^i$, respectively. Under the null hypothesis, there is no DM signal and the background best-fit parameters $(\hat{A}_{\rm bkg},\hat{\gamma}_{\rm bkg})$ are determined by maximizing the likelihood of Eq.~\eqref{eq:like_fermi} with ${\tilde{{{\cal{C}}}}_{{\cal B},{\cal W}}}=0$. In this way, it is possible to infer upper limits on the $({\tilde{{{\cal{C}}}}_{{\cal B}}},{\tilde{{{\cal{C}}}}_{{\cal W}}})$ coefficients by introducing the test statistics 
\begin{equation}
\text{TS}({\tilde{{{\cal{C}}}}_{{\cal B}}},{\tilde{{{\cal{C}}}}_{{\cal W}}}) \equiv 2\log{\left[\frac{\mathcal{L}(0,0,\hat{A}_{\rm bkg},\hat{\gamma}_{\rm bkg})}{\mathcal{L}({\tilde{{{\cal{C}}}}_{{\cal B}}},{\tilde{{{\cal{C}}}}_{{\cal W}}},\hat{A}_{\rm bkg},\hat{\gamma}_{\rm bkg})}\right]},
\end{equation}
which follows an asymptotic $\chi^2$ distribution with one degree of freedom. %\footnote{Although  we have two free parameters, ${\tilde{{{\cal{C}}}}_{{\cal B},{\cal W}}}$, we actually have only one \textit{physical} parameter, i.e. the total annihilation cross-section. Consequently, $\text{TS}({\tilde{{{\cal{C}}}}_{{\cal B}}},{\tilde{{{\cal{C}}}}_{{\cal W}}})$ is distributed as a $\chi^2$ with one degree of freedom rather than two.}
As a consequence, we can use Wilks' theorem to compute the one-sided 95\% CL upper-limit on the $({\tilde{{{\cal{C}}}}_{{\cal B}}},{\tilde{{{\cal{C}}}}_{{\cal W}}})$ coefficients by imposing $\text{TS}({\tilde{{{\cal{C}}}}_{{\cal B}}},{\tilde{{{\cal{C}}}}_{{\cal W}}})\simeq 2.71$. This condition imposes a constraint on the $({\tilde{{{\cal{C}}}}_{{\cal B}}},{\tilde{{{\cal{C}}}}_{{\cal W}}})$ plane for a given DM mass, where the allowed region is inside the contour.

\begin{figure}[t!]
    \centering
    \includegraphics[width=0.45\linewidth]{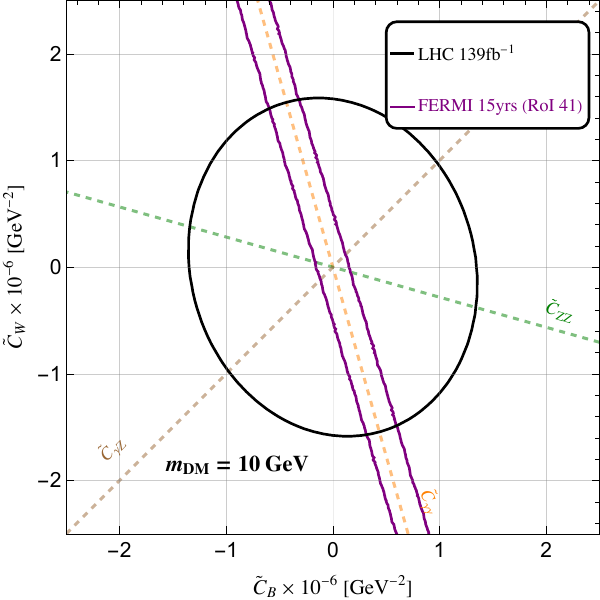}\hspace{0.5cm}
    \includegraphics[width=0.45\linewidth]{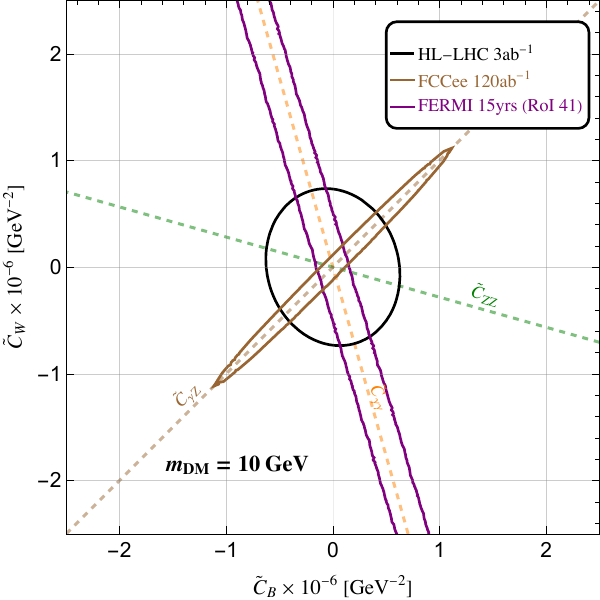} \\
\caption{Constraints in the $({\tilde {\cal C}}_{\cal B},{\tilde {\cal C}}_{\cal W})$ plane for FERMI experiment and DY processes at LHC (\textit{left panel}), HL-LHC and FCCee (\textit{right panel}). In the figures we fixed the DM mass to $m_{\rm DM}=10$ GeV. The dashed lines represent the directions corresponding to the combinations of ${\tilde {\cal C}}_{\cal B}, {\tilde {\cal C}}_{\cal W}$ that cause the couplings in Eq.~\eqref{eq:EWSB-coeff} to vanish. From the left panel, it is evident that FERMI cannot be sensitive in the direction where ${\tilde{\cal C}}_{\gamma\gamma}$ vanishes, but for small DM masses, it provides a more stringent bound in its orthogonal direction. In the right panel, we show the interplay between FERMI and \textit{near} next-future colliders, HL-LHC and FCC-ee.}
\label{fig:fermi_ell} 
\end{figure}

In Fig.~\ref{fig:fermi_ell} we compare the contour of FERMI (solid purple line) with the LHC constraint (solid black line) on the mono-$\gamma$ searches computed in Sec.~\ref{sec:LHC_DY} for a DM mass $m_{\rm DM}=10$ GeV. The relevant feature to note is that FERMI provides a stronger constraint than LHC, except along the direction  ${\tilde{{{\cal{C}}}}_{{\cal W}}}=-\text{tan}_w^{-2}\,{\tilde{{{\cal{C}}}}_{{\cal B}}}$, where the coupling to photons ${\tilde{{{\cal{C}}}}_{{\gamma \gamma}}}$ is set to zero. This happens because for DM masses below $m_Z/2$ the only kinematically available channel is into $\gamma\gamma$, and aligning towards the ${\tilde{{{\cal{C}}}}_{{\cal W}}}=-\text{tan}_w^{-2}\,{\tilde{{{\cal{C}}}}_{{\cal B}}}$ direction would completely suppress the photon flux, thus eliminating the ID constraints.

In Fig.~\ref{fig:DDandIDbound} we present the ID constraints considering a DM masses in the 10 GeV–300 GeV range. The analysis focuses on two cases: $\tilde{{\cal{C}}}_{{\cal B}} = 0$ (left panel) and $\tilde{{\cal{C}}}_{{\cal W}} = 0$ (right panel), with the constraints on the scale $\tilde{{\cal{C}}}_{{\cal B, \cal W}}^{-1/2}$. The plot differentiates between single-line and double-line search results. Single-line constraints, which consider only the $\langle \sigma v \rangle_{\gamma \gamma}$ term in Eq.~\eqref{eq:flux}, are shown as a solid purple line. These limits on the excluded cross-section $\langle \sigma v \rangle_{\gamma \gamma}$ are in excellent agreement with recent model-independent constraints obtained from 15 years of Fermi-LAT data~\cite{DeLaTorreLuque:2023fyg}. On the other hand, double-line constraints, indicated by a dashed purple line, depend on the relative signal strength between the $\gamma \gamma$ and $\gamma Z$ channels. For Rayleigh DM interactions, this ratio remains approximately constant for $m_{\rm DM} \gg m_Z/2$, and is given by $ 
\langle \sigma v \rangle_{\gamma Z}/\langle \sigma v \rangle_{\gamma \gamma} \approx  s_{2w}^2 (\tilde{{\cal{C}}}_{{\cal B}} - \tilde{{\cal{C}}}_{{\cal W}})^2/2\,(c_w^2 \tilde{{\cal{C}}}_{{\cal B}} + s_w^2 \tilde{{\cal{C}}}_{{\cal W}})^2
$, see Eq.~\eqref{eq:ann_xsecs}. This behavior implies that double-line constraints have much greater sensitivity than single-line constraints when $\tilde{{\cal{C}}}_{{\cal B}} = 0$ (see left panel of Fig.~\ref{fig:DDandIDbound}), where $\langle \sigma v \rangle_{\gamma Z}/\langle \sigma v \rangle_{\gamma \gamma} \sim O(1)$. In contrast, when $\tilde{{\cal{C}}}_{{\cal W}} = 0$, the ratio is smaller, approximately $\langle \sigma v \rangle_{\gamma Z}/\langle \sigma v \rangle_{\gamma \gamma} \sim O(10^{-1})$, so that the single and double-line constraints yields comparable limits (see right panel of Fig.~\ref{fig:DDandIDbound}). 

Finally, we compare the constraints from FERMI with those from colliders discussed in the previous sections, as illustrated in Fig.~\ref{fig:fig_sc_finalplot}. In the relevant mass range, FERMI provides tighter constraints than both the current and projected bounds from the LHC, as well as the limits from DD experiments. However, future colliders like the FCC-hh and the $\mu$C are expected to achieve stronger constraints than FERMI.

\subsection{Future Cherenkov Telescope Array Constraints}
The Cherenkov Telescope Array Observatory (CTAO) is a next-generation ground-based gamma-ray telescope~\cite{CTAConsortium:2010umy}, which will be able to measure gamma-ray signals from energy $\sim$100 GeV up to several tens of TeV with unprecedented angular and energy resolution. The latter feature will possibly allow to discriminate well-localized spectral features in the gamma-ray flux which can possibly be produced by DM annihilation.

In the specific case of DM annihilation producing gamma-ray lines, at energies relevant to CTA, where $E_\gamma \gg m_Z$, the resulting spectral line from Rayleigh DM is simply a superposition of the contributions from the $\gamma\gamma$ and $\gamma Z$ annihilation channels, with no threshold effects present.

Ref.~\cite{CTAO:2024wvb} presents the sensitivity projections for two different targets: \textit{i)} the inner 4$^\circ$ of the Galactic Center (GC) and \textit{ii)} a collection of Dwarf Spheroidal Galaxies (dSphs). To this respect, dSphs offer a more robust sensitivity than the former target given the uncertainty on the DM density profile in the GC. 

\begin{figure}[t!]
    \centering
    \includegraphics[width=0.45\linewidth]{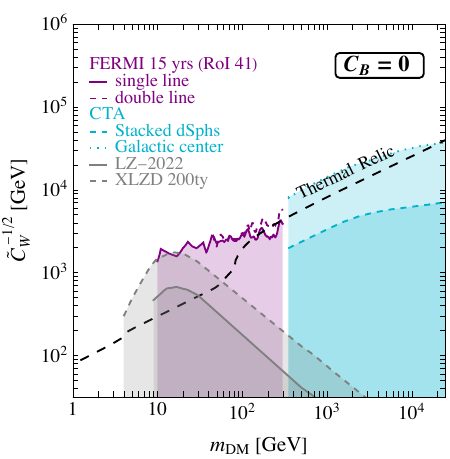}\hspace{0.5cm}
    \includegraphics[width=0.45\linewidth]{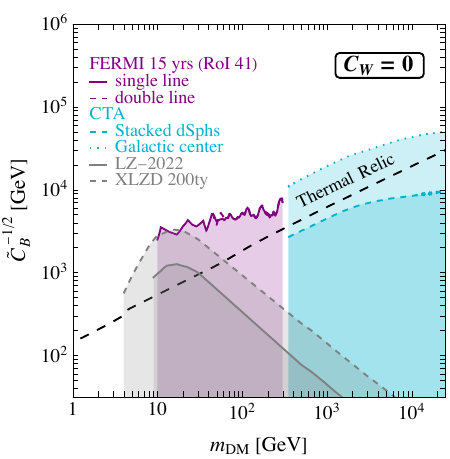} \\
\caption{ID and DD bounds for ${\tilde {\cal C}}^{-1/2}_{\cal W,B}$ as a function of DM mass for ${\tilde {\cal C}}_{\cal B} = 0$ (\textit{Left panel}) and ${\tilde {\cal C}}_{\cal W} = 0$ (\textit{Right panel}). The gray solid line corresponds to the current bound from LZ-2022, while the gray dashed line represents the future projection from a 200 tonne-year XLZD experiment. The purple solid and dashed curves denote the single-line and double-line search obtained from 15 years of FERMI data, while the dashed and dotted blue lines correspond to the limits from CTA obtained by the observations of Dwarf Spheroidal Galaxies and the Galactic Center, see the main text for more details. Rayleigh DM remains elusive for DD probes, and the current ID constraints from FERMI data are already more stringent compared to the future projections from the XLZD experiment.}

    \label{fig:DDandIDbound}
\end{figure}

For these targets, we use the 95 \% CL upper limits of Ref.~\cite{CTAO:2024wvb} obtained by considering 100 h observation of stacked dSphs both in the southern and northern hemisphere as well as the upper limits from 500 h observation of the GC by assuming an Einasto density profile~\cite{Einasto1965OnTC}. 
The results are illustrated in Fig.~\ref{fig:fig_sc_finalplot} and in Fig~\ref{fig:DDandIDbound}, where we denote the CTA limits by a light blue dashed and dotted lines for dSphs and GC observations, respectively. While these constraints offer a complementary limit to FERMI, extending to masses beyond the TeV range, it is important to note that the conservative limit coming from stacked dSphs observations falls almost entirely within the region where the EFT framework breaks down. Contrarily, the stringent constraints arising from the GC will provide useful insights within the EFT-validity region complementing the sensitivity of future collider searches in the multi-TeV range.

\section{Results }\label{sec:results}
In this work, we have presented key findings regarding the Rayleigh DM scenario, exploring current and future probes from collider physics, indirect detection, and direct detection experiments.

The HL-LHC phase is expected to provide only modest improvements over the current LHC constraints, enhancing them by almost a factor $\sim 2$. In contrast, the next-generation collider FCC-ee is projected to set significantly stronger bounds, improving by a factor of $\sim 6$ compared to the LHC. However, these improvements are primarily effective for low DM masses. Looking further ahead, colliders such as a $\mu$C and FCC-hh will provide substantial advancements, improving constraints by a factor of $\sim 10$ compared to the LHC and extending the reach to much higher energy scales and DM masses.

Among the current probes, the FERMI experiment provides the most stringent constraints, exceeding the LHC limits by less than a factor of $\sim 10$. Future projections for the CTA are highly promising, particularly in the context of galactic center analysis. CTA will cover higher DM masses and will be competitive with, or even stronger than, the next-generation colliders. For example, GC observations are expected to improve current constraints by more than a factor of $\sim 5$ compared to FCC-hh and $\mu$C. On the other hand, focusing on dSphs would yield a limit approximately five times weaker than searches directed toward the GC, resulting in a constraint that falls almost entirely within the region where the EFT framework breaks down.

Due to the elusive nature of the Rayleigh DM scenario for direct detection, the current bounds from LZ-2022 are weaker than those provided by other probes. Additionally, future projections for experiments like XLZD are expected to remain subdominant, offering weaker constraints than those already set by FERMI data.

\section{Conclusions}\label{sec:concl}
We have presented  current and expected sensitivity on the Rayleigh Dark Matter scenario, 
focusing on both collider physics and indirect and direct Dark Matter detection experiments. We find that HL-LHC  will slightly improve the current LHC constraints. 
In contrast, a high luminosity phase at an $e^+e^-$ machine, e.g the $Z$-pole run of the FCC-ee or CEPC is going to be particularly sensitive for low Dark Matter masses and can improve HL-LHC results by one order of magnitude on the scale that suppresses the contact interactions. 
Current limits and future projections from direct and indirect detection experiments are expected to remain competitive, potentially surpassing those from FCC-ee and HL-LHC.

Looking ahead, colliders such as the FCC-hh and a 10 TeV muon collider will probe further higher energy scales on an even wider Dark Matter mass range. We have shown in Appendix~\ref{sec:app_fcc} for the concrete case of FCC-hh  that the reach does not  crucially  depend on presently unknown detector acceptance and geometry. Therefore we conclude that these projects further in the future have excellent prospect to probe Rayleigh Dark Matter.  

Future indirect detection experiments like CTA are also expected to impose stringent constraints on Rayleigh Dark Matter in the multi-TeV range, in synergy with the capabilities of upcoming colliders. 
The projected sensitivity of the future direct detection experiment XLZD will enable probing of thermal Dark Matter up to 100 GeV. Furthermore, it will provide complementary insights to a potential signal at the LHC, confirming its cosmological origin.
We studied in detail the issue of the validity of the EFT description at colliders, finding that this is a point on which higher energy machines, such as LHC, FCC-hh and muon colliders need to pay attention. The EFT bounds can sometimes  correspond to new physics particles at a mass scale accessible at the same machine, thus it is important to study these models of Dark Matter keeping in mind possible UV completions.
In Appendix~\ref{app:UV} we discuss the details of possible UV completions of Rayleigh Dark Matter, highlighting the range of applicability of the obtained results and the possible interplay with direct searches at colliders.

\medskip
To summarize, taking as a benchmark the thermally produced Rayleigh Dark Matter scenario, we find that, thanks to the complementarity between collider experiments and cosmological probes, Dark Matter at the hundreds of GeV scale can be thoroughly tested with the next generation of experiments. For lighter candidates, upcoming forecasts will explore uncharted parameter space, significantly surpassing the thermal Dark Matter benchmark. The expected results will therefore probe both minimal thermal histories for WIMP-like candidates, but also other less minimal scenarios corresponding to weaker coupling.

\section*{Acknowledgments}
We thank Andrea Wulzer for sharing useful information regarding the background analysis at muon collider.
We acknowledge Pedro De la Torre Luque for sharing the useful discussion on the FERMI dataset. We appreciate valuable 
discussion with Luca Vecchi about the realization of the tree level UV completion. We are grateful to Robert Ziegler for very useful discussions concerning the matching of the loop induced UV completion.
DBu, AD, RF, GM and PP are supported in part by the European Union - Next
Generation EU through the MUR PRIN2022 Grant n.202289JEW4.
DBa acknowledges the use of the LXPLUS High Performance Computing facility at CERN in the completion of this study.

\appendix

\section{UV completions}\label{app:UV}
In this Appendix, we explore two benchmark scenarios for UV completions: loop-level and tree-level frameworks. In the first case, new fermions are introduced via Yukawa-like interactions with the Dark Matter (DM) field. For the second scenario, we investigate a massive spin-2 resonance that provides a tree-level UV completion for the Rayleigh EFT discussed in the main text. The main goal of this section is to provide examples, as a proof of principle, illustrating how the analyzed EFT can be UV completed. The main goal of this section is to provide examples illustrating how in principle the analyzed EFT can be UV completed, what matching conditions are to be expected and what relations it is reasonable to assume between the effective scales $\Lambda$ of the contact interactions and the physical masses of the new degrees of freedom in the UV completion.
\subsection{Loop induced - Yukawa model}

A simple UV framework that generates the Rayleigh operator at the one-loop level involves a Yukawa coupling between the scalar DM field $\phi$ and two fermions. This UV completion can be realized in two distinct ways.  The first way, known as the \textit{lepton portal model}~\cite{Lepton-portal-model} and referred to here as \textit{scenario} I, involves a SM lepton and a BSM fermion. In contrast, the second approach, labeled as \textit{scenario} II, features two different BSM fermions.

\textbf{\textit{Scenario} I:} Here the interaction lagrangian is defined as
\begin{equation}
    \mathcal{L} = \lambda \, \phi \, \bar{\psi}_L \ell_R + \text{h.c.}
\end{equation}
where \(\phi\) is the scalar DM field, a singlet under \(SU(2)_L \times U(1)_Y\), \(\ell_R\) is a right-handed SM lepton, and \(\psi\) is a BSM fermion singlet under $SU(2)_L$ with hypercharge \(Y_\psi = -1\). Following~\cite{Ibarra,Giacchino:2014moa}, we compute the annihilation amplitude for the \(\phi\phi \rightarrow \gamma\gamma\) process at one loop and in the limit of zero relative velocity, considering only \(s\)-wave terms. At this point we perform the matching between the UV amplitude and the EFT Rayleigh one
\begin{equation}
    \mathcal{A}_{\text{Rayleigh}}=-i\frac{\alpha}{\pi}{\cal C}_{\gamma\gamma}\bigg[(k_1\cdot k_2)(\epsilon(k_1)\cdot\epsilon(k_2))-(k_1\cdot\epsilon(k_2))(k_2\cdot\epsilon(k_1)) \bigg].
    \label{eq:ampeftray}
\end{equation}
Here $\epsilon$ represents the photon polarization vector and $k_{1,2}$ denote the momenta of the final state photons. The relation between the Wilson coefficient \({\cal C}_{\gamma\gamma}\) and the UV coupling \(\lambda\), in the $m_\psi \gg m_\text{DM}$ limit, is given by:
\begin{equation}
    C_{\gamma\gamma} = \frac{2\lambda^2}{3 m_\psi^2},
\end{equation}
where \(m_\psi\) is the mass of the BSM lepton. According to the low-energy theorem~\cite{Shifman:1979eb}, once both leptons are also integrated out, 
the contributions from heavy and light fermions cancel each other, leading to a Wilson coefficient that vanishes at zeroth order in the \(m_{\rm DM}/m_\psi\) expansion~\footnote{We  have also verified this point by an explicit calculation through the code Matchete~\cite{Fuentes-Martin:2022jrf} and by using the expressions in Refs.~\cite{Ibarra,Giacchino:2014moa}, see our result in Eq.~\eqref{eq:match_scenarioII}.}.  The perturbative unitarity bounds on the Yukawa coupling \(\lambda\) can be derived using the method presented in~\cite{Allwicher:2021rtd} and for a real scalar field read as
\begin{equation}
    \lambda_{\rm} \leq\lambda_{\rm max}= 2\sqrt{\frac{2\pi}{3}} \approx 2.9.
    \label{eq:pertUnitUVcoupling}
\end{equation}

This scenario represents somehow a minimal UV benchmark, with the peculiarity that it allows interactions involving the DM and a combination of the three SM leptons, \(\ell = e, \mu, \tau\). As a result, for energies above the lightest SM lepton, other operators are expected to appear in addition to the Rayleigh EFT defined in~\eqref{eq:ops_1}. For instance the operator $\phi^2 \bar{\ell}\slashed{D}\ell\to m_\ell \, \phi^2 \bar{\ell}\ell$, as already discussed in~\cite{Kavanagh:2018xeh}.
In this case, a detailed analysis of this specific UV completion is required to take into account possible further signals generated by the interactions beyond the Rayleigh one. We do not wish to discuss these UV specific details, as it  is possible to provide other scenarios where these  effects are absent. One such example is discussed in the following as a proof of existence. 

\textbf{\textit{Scenario} II:} In the second UV scenario we introduce two BSM fermions ($\psi,F$), both singlets under $SU(2)_L$, with equal hyperchanges, $Y_\psi=Y_F$, namely
\begin{equation}
    \lag=\tilde{\lambda}\phi\bar{\psi}_L F_R+h.c. \ .
    \label{eq:lagLFUV}
\end{equation}
 The mass hierarchy is assumed to satisfy \(m_\phi \ll m_F \lesssim m_\psi\). The annihilation amplitude is computed as in Scenario I and the general expression is expanded in the limit of a small mass splitting between the two heavy BSM fermions, \(\epsilon = (m_\psi^2 - m_F^2)/m_{\rm DM}^2\), and subsequently, we take the small $m_{\rm DM}$ limit. Under these assumptions, we perform the one-loop matching between the Wilson coefficient \({\cal C}_{\gamma\gamma}\) and the UV coupling \(\tilde{\lambda}\). At the zeroth order in the mass splitting $\epsilon$ the Wilson coefficient reads
\begin{equation}\label{eq:match_scenarioII}
    {\cal C}_{\gamma\gamma} = \frac{11\tilde{\lambda}^2\, Y_{\psi}^2}{90} \frac{m_{\rm DM}^2}{m_\psi^4}.
\end{equation}
As a result, the Rayleigh operator is generated even in the degenerate case \(m_\psi = m_F\). This allows us to consider this particular case as a viable scenario to evade constraints from EW searches at the LHC. The unitary bound on the Yukawa $\tilde \lambda$ is the same as before ($\tilde \lambda\lesssim2.9$), depending only on the EW quantum numbers of the involved fields.

\begin{figure}[t!]
\centering
    \includegraphics[width=0.45\linewidth]{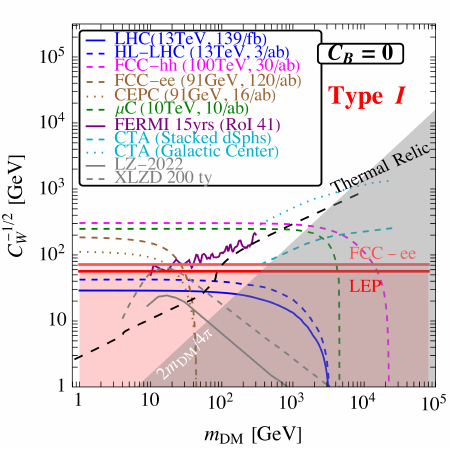}\hspace{0.5cm}
    \includegraphics[width=0.45\linewidth]{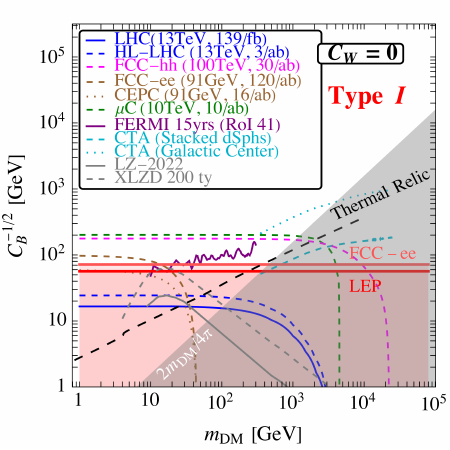}
    \includegraphics[width=0.45\linewidth]{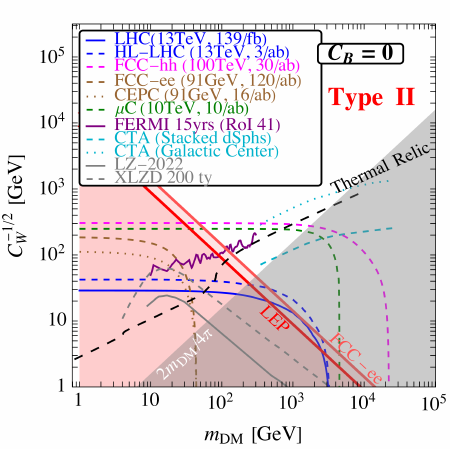}\hspace{0.5cm}
    \includegraphics[width=0.45\linewidth]{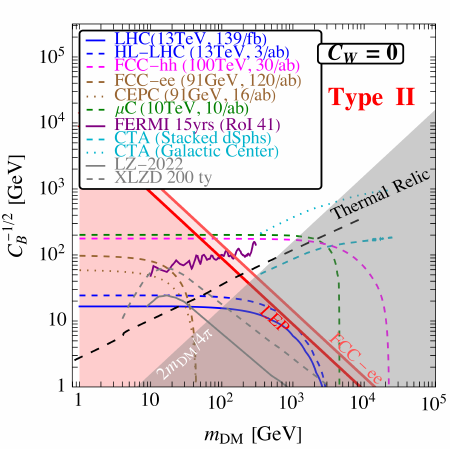}
    \vspace{0.5cm} 
    \caption{Bounds on the loop normalized effective scale \({\cal C}_{\gamma\gamma}^{-1/2}\) as a function of the DM mass in the case of \(\mathcal{C}_B = 0\) (\textit{Left panels}) and \(\mathcal{C}_W = 0\) (\textit{Right panels}), for both type I (top row) and type II (bottom row) Yukawa-like UV scenarios. The legend is the same as in Fig.~\ref{fig:fig_sc_finalplot}. Direct searches for EW final states from LEP (red) and FCC-ee (light-red) are also shown fixing $Y_\psi=1$ and $\tilde \lambda\sim\tilde \lambda_{\rm max}$. As a result of Eq.~\eqref{eq:DiffScenarios}, in the type I scenario there is no dependence on the DM mass in the EW direct probes constraints. In contrast, for the type II scenario, the effective loop scale is related to the mass of the heavy fermion, \(m_\psi \sim m_F\), with a dependence proportional to \(m_{\rm DM}^{-1}\).
}
    \label{fig:fig_wc_finalplot}
\end{figure}

\smallskip
We emphasize that the bounds obtained in the previous sections in EFT and shown in Fig.~\ref{fig:fig_sc_finalplot} must be rescaled by a loop factor for this UV-model. In Fig.~\ref{fig:fig_wc_finalplot}, we plot the bound for \(\cal C_{\gamma\gamma}\) in place of the the plots for  \(\cal \tilde{C}_{\gamma\gamma}\) of Fig~\ref{fig:fig_sc_finalplot}. 

Notice that, since in these UV completions new fermions are expected to appear at the EW scale, 
we need to take into account LEP direct searches for  new EW matter, which exclude new electroweak matter up to \(m_\psi \lesssim 95 \, \text{GeV}\). Similarly  the anticipated sensitivity of the FCC-ee or CEPC is expected to strengthen the exclusion limit to \(m_\psi \lesssim 120 \, \text{GeV}\) at the $Zh$ threshold run. For illustration, the LEP direct search constraint on the heavy lepton mass translates into a limit on the effective scale \({\cal C}^{-1/2}_{\gamma\gamma}\) as follows:
\begin{equation}
    \begin{cases}
        {\cal C}^{-1/2}_{\gamma\gamma} = \displaystyle\sqrt{\dfrac{3}{2}}\dfrac{m_\psi}{\lambda  }\bigg|_{\lambda_{\rm max}} \gtrsim 40\, \text{GeV}, & \quad \text{(Scenario I)} \\\\
        {\cal C}^{-1/2}_{\gamma\gamma} = 
        \displaystyle\sqrt{\frac{90}{11}} 
        \dfrac{1}{\tilde{\lambda}\,|Y_{\psi}|} \dfrac{m_{\psi}^2}{m_{\rm DM}} \bigg|_{\tilde{\lambda}_{\text{max}}} \gtrsim 89\,\text{GeV}\,\left(\dfrac{100\, \text{GeV}}{m_{\text{DM}}}\right), & \quad \text{(Scenario II)} 
        \label{eq:DiffScenarios}
    \end{cases}
\end{equation}
where the UV couplings \(\lambda\) and \(\tilde{\lambda}\) are pushed to their perturbative unitarity limits, as defined in Eq.\,\eqref{eq:pertUnitUVcoupling}, and $|Y_\psi|$ is chosen to be one in both scenarios. More stringent constraints can be expected from running FCC-ee or CEPC at energies higher than LEP. These bounds at these future colliders are expected to be linearly increasing with the center of mass energy with respect to the LEP ones. 
As a result, the constraint on \({\cal C}^{-1/2}_{\gamma\gamma}\) is independent of the DM mass in the type I scenario, while it scales as \(m_{\rm DM}^{-1}\) in the type II scenario. In Fig.~\ref{fig:fig_wc_finalplot} we show the direct bounds from LEP and FCC-ee or CEPC, assuming 240 GeV center of mass energy. These limits are represented by the shaded red and light-red regions, respectively.
We want to stress that LEP search provides a more stringent constraint compared to current and projected bounds from the LHC, HL-LHC, FERMI, and both current and future direct detection experiments. The FCC-ee and CEPC, and especially a future muon collider or FCC-hh, could offer additional insights into this benchmark scenario. Our conclusion is that, focusing on the loop-level UV completion, near future experiments at colliders and direct detection experiments are unlikely to significantly strengthen the constraints on these particular UV scenarios. Nevertheless, further experiments of the indirect detection type and further future colliders such as FCC-hh and muon colliders can prove new regions of the phase-space of these UV scenarios.  

Additionally, since new BSM fermions are charged under $U(1)_Y$, deviations in the EW observables are expected. These observables are tightly constrained by LEP data~\cite{Barbieri:2004qk}. Specifically, the contribution to the \(\widehat{Y}\) parameter of a SU(2) singlet  is given by~\cite{Cirelli:2005uq, Bottaro:2022one}
\begin{equation}\label{eq:WY_par}
    \begin{split}
        %&\widehat{W} = \frac{\alpha_{\rm em} \, \cot^2 \theta_{\rm w}}{180 \pi} \frac{m_Z^2}{m_F^2} \, \kappa \, n (n^2 - 1) \simeq 3.8 \, \kappa \times 10^{-7} \left( \frac{1 \, \text{TeV}}{m_F} \right)^2 n (n^2 - 1) \, , \\
        &\widehat{Y} = \frac{\alpha_{\rm em}}{15 \pi} \frac{m_Z^2}{m_\psi^2} \, \kappa \, \, Y^2 \simeq 3.4 \, \kappa \times 10^{-5} \left( \frac{200 \, \text{GeV}}{m_\psi} \right)^2 Y_\psi^2 \, ,
    \end{split}
\end{equation}
where $\kappa=1$ for a Dirac fermion, $Y$ is the representation of the fermion under $U(1)_Y$ gauge group.
From Eq.~\eqref{eq:WY_par}, we can derive constraints on \(m_{\rm \psi}\) by applying the limits set by LEP data, $\hat{Y}=4\times10^{-4}$, as asserted in~\cite{Falkowski:2015krw, Farina:2016rws}. We obtain that for $Y_\psi\lesssim3$ the required mass of the heavy fermion is comparable to EW scale $v/\sqrt{2}\sim174\,$GeV. As a viable example, $Y_\psi=4$ requires $m_\psi\gtrsim233\,$GeV giving a more stringent constraint with respect to LEP bound.

\subsection{Tree-level Spin 2 contribution}
As discussed in ~\cite{Craig_2020}, Rayleigh operators can be UV-completed at tree-level by integrating out a spin-2 particle that couples to the EW gauge bosons.  
Now, as it is known, spin-2 interactions  are not renormalisable by itself and requires further dynamics 
emerging at scales higher than those where the Rayleigh operators are generated. 
Achieving an ultraviolet completion of this spin-2 EFT is beyond the purpose of our work. It requires going beyond four-dimensional quantum field theory,  by incorporating for example  gravitational theories with extra dimensions, such as  ADD~\cite{Arkani_Hamed_1999} and Randall-Sundrum models~\cite{Randall_1999}. In these models, spin-2 interactions to matter arise by linearizing the metric around flat space, leading to standard couplings to the energy-momentum tensor.  All in all, the EFT of spin-2 particles reads 
\begin{equation}
    \mathcal{L} \supset -\frac{1}{2\Lambda_{\mathrm{R}}} R^{\mu\nu} \left[ c_{\gamma} T_{\mu\nu}^{(\rm EM)} + c_{\phi} T_{\mu\nu}^{(\phi)} \right],
    \label{eq:spin2EFT}
\end{equation}
where \( R^{\mu\nu} \) is the  symmetric tensor of the spin-2 resonance~\footnote{For spin-1 particle,  the  tensor field  \( R^{\mu\nu} \) would be antisymmetric and thus the interaction in Eq.~\eqref{eq:spin2EFT} would vanish. In the case of spin-0 particle, the tensor would be of rank zero, i.e  \( R^{\mu\nu} =g^{\mu\nu} R\) and   again  the interaction would vanish being $T_{\mu\nu}^{(\rm EM)}$ traceless}, and \( T_{\mu\nu}^{(\rm EM)} \) and \( T_{\mu\nu}^{(\phi)} \) are the energy-momentum tensors of the electromagnetic and DM fields, respectively. The couplings $c_{\gamma,\phi}/\Lambda_R$ are ultimately related to the Planck scale and the size of the extra-dimension.  It is important to note that, unlike the graviton, the coupling of the spin-2 with DM and photon is not universal, as the coefficients \(c_{\gamma}\) and \(c_{\phi}\) can take different values. Moreover, the fact that the spin-2 resonance does not necessarily couple universally to SM particles is also crucial since the gluon coupling  \(c_G\) can be set to zero, thus avoiding the tree-level generation of a DM-gluon coupling below \(\Lambda_{\rm tree}\). Achieving such a coupling structure may not be trivial in a concrete model~\cite{Agashe:2016uq}.

\begin{figure}[ht!]
    \centering
    \includegraphics[width=0.5\linewidth]{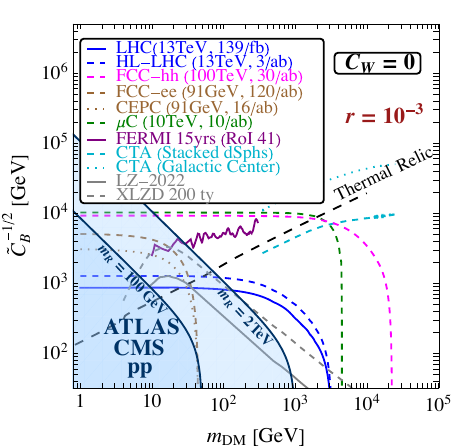}
    \caption{${\tilde{\cal C}_{\cal B}^{-1/2}}$ as a function of the DM mass fixing ${\tilde {\cal C}}_{\cal W}=0$. Similarly to Fig.~\ref{fig:fig_sc_finalplot} we show constraints arising from collider physics, as well as current and future direct and indirect detection experiments. Additionally we compared show limits from the search for a spin-2 particle at LHC. The latter is analyzed for two different masses of the spin-2 resonance, specifically for $m_{\rm R} = 100\,\text{GeV}$ and $2\,\text{TeV}$,  with ATLAS and CMS pp searches. We consider a benchmark for the branching ratio of the decay channel into two photons  $r = 10^{-3}$.}
    \label{fig:plspin2}
\end{figure}
\begin{figure}[ht!]
    \centering
    \includegraphics[width=0.8\linewidth]{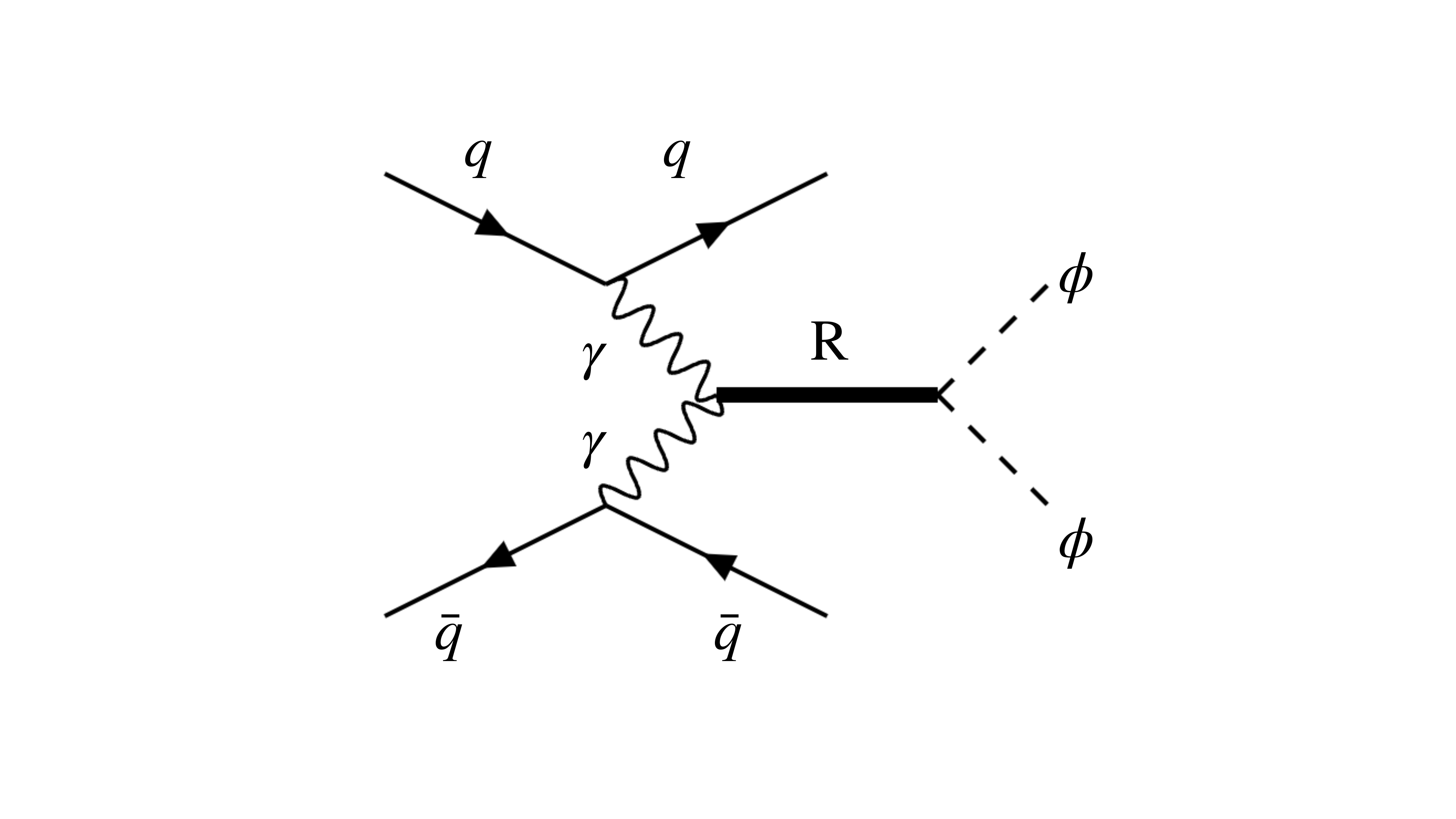}
    \caption{
Representative Feynman diagram for neutral VBS pair-production of scalar $\phi$ particles through the exchange of a spin 2 particle $\rm R$ at LHC.}
\label{fig:vbf_spin2}
\end{figure}

To calculate ${\tilde {\cal C}}^{-1/2}_{\gamma\gamma}$, we match the UV theory of Eq.~\eqref{eq:spin2EFT} to the Rayleigh operator 
by computing the amplitude    of the process $\gamma\gamma\rightarrow R\rightarrow \phi\phi\,.$
Using the Feynman rules given in~\cite{Han_1999}, the amplitude in the full theory is
\begin{equation}
    \mathcal{A}_{\text{R}}=-\frac{4 ic_\gamma c_\phi}{\Lambda_{\rm R}^2}\bigg(\frac{m_{\rm DM}^2}{M_{\rm R}^2}\bigg)\bigg[(k_1\cdot k_2)(\epsilon(k_1)\cdot\epsilon(k_2))-(k_1\cdot\epsilon(k_2))(k_2\cdot\epsilon(k_1)) \bigg],
\end{equation}
whereas the amplitude of the effective Rayleigh operator reads as in Eq.~\eqref{eq:ampeftray} rescaled by a factor $(\alpha/8 \pi)^{-1}$ (see Eq.~\eqref{eq:scales}) since in this particular benchmark UV completion the spin-2-photon coupling is not generated through the minimal electroweak coupling.
Now the expression of ${\tilde {\cal C}}^{-1/2}_{\gamma\gamma}$ in terms of $\Lambda_{\rm R}$ is
\begin{equation}\label{eq:match}
    {\tilde {\cal C}}^{-1/2}_{\gamma\gamma}=\frac{\sqrt{2}}{\sqrt{c_\gamma c_\phi}}\bigg(\frac{m_{\rm R}}{m_{\rm DM}}\bigg)\Lambda_{\rm R}
\end{equation}
Before imposing the constraints on ${\tilde {\cal C}}^{-1/2}_{\gamma\gamma}$ from spin-2 searches, it is useful to rewrite Eq.~\eqref{eq:match} in terms of the spin-2 branching ratio. To this purpose, we define the effective coupling
\begin{equation}\label{eq:gR}
    g_{\rm R\gamma} = \frac{c_\gamma}{2 \Lambda_{\rm R}}
\end{equation}
and compute the decay widths for both decay channels into photons and DM:
\begin{equation}
\Gamma(R \rightarrow \gamma \gamma) = g_{\rm R\gamma}^2\frac{m_{\rm R}^3}{40 \pi}
\,,\quad\Gamma(R \rightarrow \phi \phi) = g_{R\gamma}^2\frac{m_R^3}{1920\pi}  \left(1 - 4 \frac{m_{\rm DM}^2}{m_R^2}\right)^{5/2},
\end{equation}
which give the following branching ratio 
%(see Eq. \eqref{eq:BR_def})
\begin{equation} \label{eq:r-ratio}
    r=  \frac{{\Gamma}_{\rm R\rightarrow\gamma\gamma}}{{\Gamma}_{\rm R\rightarrow \gamma\gamma} + {\Gamma}_{\rm R\rightarrow \phi\phi}}=       
    \frac{1}{1 + \frac{c_\phi^2}{c_\gamma^2} \, \tilde{f}(m_R, m_{\rm DM})}
\end{equation}
with \( \tilde{f}(m_R, m_{\rm DM}) \equiv (1/48) \left(1 - 4 m_{\rm DM}^2/m_{\rm R}^2\right)^{5/2} \). The latter equation gives a relation among the $c_{\gamma}$ and $c_{\phi}$ couplings at fixed branching $r$
\begin{equation}\label{eq:c_rel}
    c_\phi=c_\gamma\sqrt{\frac{1-r}{r}\frac{1}{\Tilde{f}(m_R,m_{\rm DM})}}.
\end{equation}
Plugging Eqs.~\eqref{eq:gR},~\eqref{eq:c_rel} in Eq.~\eqref{eq:match}, we write ${\tilde {\cal C}}^{-1/2}_{\gamma\gamma}$ as
\begin{equation}\label{eq:final_spin2_match}
{\tilde {\cal C}}^{-1/2}_{\gamma\gamma}=\frac{1}{\sqrt{2}}\bigg(\frac{r\Tilde{f}(m_{\rm R},m_{\rm DM})}{1-r}\bigg)^{1/4}\frac{M_{R}}{m_{\rm DM}}\frac{1}{g_{\rm R\gamma}}
\end{equation}
The limits on the $g_{R\gamma}$ coupling are computed in Ref.~\cite{dEnterria:2023npy}. These constraints are derived from the latest ATLAS and CMS measurements of exclusive \(\gamma\gamma\) production in Pb-Pb and $pp$ ultra-peripheral collisions (UPCs)~\cite{CMS:2022diphoton, 2023, Tumasyan_2022, Aad_2019, Sirunyan_2019} through the diagrams in Fig.~\ref{fig:vbf_spin2}. 
Although the current data primarily target axion-like particles (ALPs), they can be reinterpreted as experimental bounds on the spin-2 photon coupling. The data provide exclusion limits on the coupling strength down to \( g_{\rm R\gamma} \approx 1 - 0.05\; \text{TeV}^{-1} \) for spin-2 masses in the range \( m_{\rm R} \approx 100\, \text{MeV} - 2\, \text{TeV} \). 
It is important to note that Ref.~\cite{dEnterria:2023npy} focuses on the  photophilic scenario ($r=1$), while if we allow a decay channel into DM $r< 1$ is implied. To take into account the value of $r$ it is possible to recast the bound by noticing that the limits on $g_{R\gamma}$ scales as $\sigma_R^{-1/2}$, with $\sigma_R$ being the cross-section for the spin-2 production (see Eq.~(18) of Ref.~\cite{dEnterria:2023npy}). Since $\sigma_R$ is proportional to $g_{R\gamma}^2 r$, Eq.~(18) of Ref.~\cite{dEnterria:2023npy} implies that $g_{R\gamma}=g_{R\gamma}^{(r=1)}  r^{-1/4}$, with the bound on $g_{R\gamma}^{(r=1)}$ provided in Fig. 5 of the same reference. Once this scaling is replaced in Eq.~\eqref{eq:final_spin2_match}, we derive the corresponding limits in the $(m_{\rm DM},\,{ \tilde {\cal C}}^{-1/2})$ planes. The results, illustrated in Fig.~\ref{fig:plspin2}, are represented by the blue shaded regions, corresponding to different benchmark values of the spin-2 mass while keeping the branching ratio $r$ fixed. Notably, for each contour, the bound on ${ \tilde {\cal C}}^{-1/2}_{\gamma\gamma}$ weakens as the DM mass approaches the threshold $m_{\rm R}/2$, beyond which the $R \to \phi\phi$ decay becomes kinematically inaccessible.

\begin{figure}[t!]
    \centering
    \includegraphics[width=0.5\linewidth]{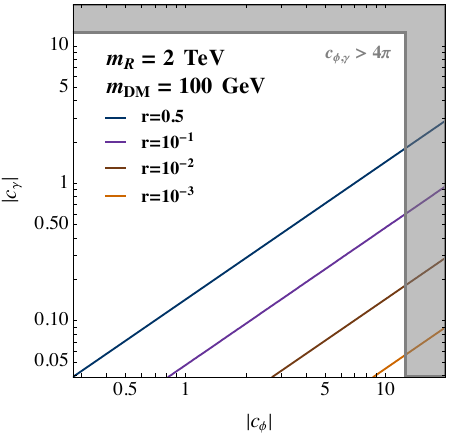}
    \caption{The $(c_\phi, c_\gamma)$ plane is shown for different branching ratios $r$, with fixed values of the spin-2 and DM masses, $m_{\rm R} = 2\,\text{TeV}$ and $m_{\rm DM} = 100\,\text{GeV}$. The gray shaded region is excluded due to perturbative unitarity constraints.}
    \label{fig:cphicgamma}
\end{figure}

Even for a significant suppression of the di-photon rate due to $r\simeq 10^{-3}$ the bounds from direct searches of the putative particle originating the Rayleigh contact interaction are quite strong.
To put in context these results, 
in Fig.~\ref{fig:cphicgamma} we show $r$ iso-lines in the plane $c_{\gamma}, c_{\phi}$
for fixed \( r, m_{\rm R} \), and \( m_{\rm DM} \) within the perturbative unitarity regime. 
As expected from  Eq.~(\ref{eq:r-ratio}), attaining a small $r$ requires hierarchical couplings $ c_{\gamma} \ll c_{\phi}$, which may require some model building to be attained.   
This puts once again the attention on a close interplay of direct searches for the UV completion generating the Rayleigh operator and the search for effects of the contact interaction itself.

\section{Modified \texorpdfstring{$p_T$}{pT} and \texorpdfstring{$\eta$}{eta} requirements for VBS processes at FCC-hh}\label{sec:app_fcc}

\begin{figure}[t!]
\begin{center}
\includegraphics[width=0.5\textwidth]{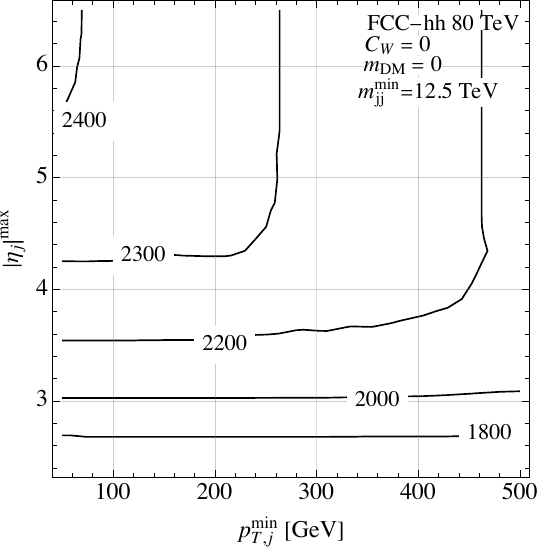}
\hfill
\end{center}
\caption{
Bound on $\tilde{\cal C}_{\tilde{\cal B}}$ as defined in Eq.\,\eqref{eq:strong_matching} for FCC-hh with $\sqrt s=80\,$TeV and an integrated luminosity of 30\,ab$^{-1}$ fixing the benchmark 
$\tilde{\cal C}_{\tilde {\cal W}}=0$ and $m_{\rm DM}=0$ in function of 
$p_{T,{\rm min}}^{j}$ and $|\eta_j|^{\rm max}$.
}
\label{fig:FCC_hh_eta_pt}
\end{figure}

In Sec.\,\ref{sec:FCC-hh_VBF} we have estimated the reach of a future hadron collider operating at $\sqrt s=80,\,100\,$TeV by considering the same selection cuts of the 13\,TeV ATLAS searches\,\cite{ATLAS:2022yvh}, with the expection of the final $m_{jj}$ binning, over which we scan in order to maximize the reach. In particular, following ATLAS, we have imposed the following selection on the jets transverse momenta and pseudo-rapidities
\be
p_T^{j_{1,2}}>80\,{\rm GeV},\,50\,{\rm GeV}~~~{\rm and}~~~|\eta_{j}|<4.5 \ .
\ee
Here we show how the obtained bounds change if one modifies the above requirements, which ultimately depend on the details of the detector that will be built for FCC-hh and that are largely unknown. For concreteness we consider the case of FCC-hh with $\sqrt s=80\,$TeV and an integrated luminosity of 30\,ab$^{-1}$, choosing the benchmark $\tilde{\cal C}_{\tilde {\cal W}}=0$ and $m_{\rm DM}=0$. We fix $m_{jj}=12.5\,$TeV as in Sec.\,\ref{sec:FCC-hh_VBF} and
in Fig.\,\ref{fig:FCC_hh_eta_pt} we show the iso-contour of the bound on $\tilde{\cal C}_{\cal B}$ in the $(p_{T,{\rm min}}^j,|\eta_j|^{\rm max})$ plane, for a simplified analysis in which both jets must have transverse momentum greater than a common threshold denoted by $p_{T,\rm min}^{j}$. With this setting we can  show the results in a two-dimension $(p_{T,{\rm min}}^j,|\eta_j|^{\rm max})$  plane. As expected, given the VBS signal production topology, the bounds become stronger at fixed $p_{T,{\rm min}}^j$ by increasing the angular acceptance of the detector $|\eta_j|^{\rm max}$ up to a value of $\sim 5$, while at fixed $|\eta_j|^{\rm max}$ one obtains the strongest bounds while relaxing the minimum transverse momentum imposed on the jets.

\newpage
%%%%%%%%%%%%%%%%%
%%%	REFERENCES	   %%%		
%%%%%%%%%%%%%%%%%

\bibliographystyle{JHEP}
{\footnotesize
\bibliography{biblio}}
\end{document}